\documentclass[a4paper,twocolumn,showpacs,pra,superscriptaddress,eqsecnum]{revtex4} 
\usepackage{graphicx}
\usepackage{dcolumn} 
\usepackage{bm} 
\usepackage{amsmath}
\usepackage{amsfonts} 
\usepackage{amssymb}
\usepackage{color}

\usepackage[normalem]{ulem}

\begin{document}
\title{Thermodynamics of the BCS-BEC crossover}

\author{R. Haussmann}
\affiliation{Fachbereich Physik, Universit\"at Konstanz, D-78457 Konstanz, Germany}

\author{W. Rantner}
\affiliation{Institut f\"ur Theoretische Physik, Universit\"at Innsbruck, Technikerstrasse 25, 
A-6020 Innsbruck, Austria}

\author{S. Cerrito}
\affiliation{Technische Universit\"at M\"unchen, James-Franck-Strasse, D-85748 Garching, Germany}
\affiliation{Institut f\"ur Theoretische Physik, Universit\"at Innsbruck, Technikerstrasse 25, 
A-6020 Innsbruck, Austria}

\author{W. Zwerger}
\affiliation{Technische Universit\"at M\"unchen, James-Franck-Strasse, D-85748 Garching, Germany}

\date{published version, 8 February 2007}

\begin{abstract}
We present a self-consistent theory for the
thermodynamics of the BCS-BEC crossover in the normal
and superfluid phase which is both conserving and gapless.
It is based on the variational many-body formalism
developed by Luttinger and Ward and by DeDominicis and Martin.
Truncating the exact functional for the entropy to that obtained within
a ladder approximation, the resulting self-consistent integral equations
for the normal and anomalous Green functions are solved numerically  for
arbitrary coupling. The critical temperature,
the equation of state and the entropy are determined as a function
of the dimensionless parameter $1/k_Fa$, which
controls the crossover from the BCS-regime of extended
pairs to the BEC-regime of tightly bound molecules.
The tightly bound pairs turn out to be described by a Popov-type
approximation for a dilute, repulsive Bose gas.
Even though our approximation does not capture the
critical behaviour near the continuous superfluid transition,
our results provide a consistent picture for the complete crossover thermodynamics
which compare well with recent numerical and field-theoretic approaches  
at the unitarity point.
\end{abstract}


\maketitle

\section{Introduction}
\label{section_1}

The problem of a two component attractive Fermi gas
near a resonance of the s-wave scattering length describing
the effective interaction is one of the basic many-body
problems which has been brought
into focus by the recent realization of molecular condensates
in ultra-cold Fermi gases \cite{Greiner,Jochim,Zwierlein1}  and the
subsequent exploration of the crossover from a Bose-Einstein-Condensate
(BEC) to a BCS-like state of weakly bound fermion pairs
\cite{crossoverexp}.  Clear signatures for the existence of
paired fermion superfluidity with cold atoms have been provided by  
spectroscopic measurements of the gap \cite{Chin} and the observation of
a vortex lattice on the BCS-side of the transition \cite{Zwierlein2}.
The ability of tuning the interaction in cold Fermi gases through
Feshbach resonances relies on
the resonant coupling of the scattering state near zero energy
of two colliding atoms with a bound state in a closed channel \cite{Julienne}.
A particularly challenging problem arises right at
the Feshbach resonance, where the two-particle scattering length is infinite
\cite{Heiselberg01,Baker99}. Precisely at this point and for broad Feshbach
resonances, where the range $r^{\star}$
of the effective interaction is much smaller than the mean interparticle
spacing \cite{Bruun,Diener,Simonucci,Burnett}, the full many-body problem has the
Fermi energy $\varepsilon_F$ as the only energy scale. As pointed out by Ho \cite{Ho},
the thermodynamics of the unitary Fermi gas is then a function only of
the dimensionless temperature $\theta=T/T_F$. More generally, as emphasized
recently by Nikolic and Sachdev \cite{Nikolic}, the universality also extends to
the behaviour away from the Feshbach resonance, as long as the broad
resonance condition $k_Fr^{\star}\ll 1$ is obeyed. Thus, for instance, the
critical temperature $T_c/T_F$ for the transition to superfluidity is a  
universal function of the inverse coupling constant $1/k_Fa$.

A quantitative theoretical understanding of the
many-body problem near a Feshbach resonance has been developed
recently through numerical calculations. In particular, at zero  
temperature and for a homogeneous system, fixed-node Green function Monte Carlo  
calculations provide quantitative results for the gap parameter \cite{Carlson03},
the equation of state \cite{Astra1},
and also the momentum distribution, the condensate fraction and the pair size
\cite{Astra2} of the ground-state for arbitrary values of $1/k_Fa$.
As expected in the case of an s-wave resonance \cite{Volovik},
these quantities all evolve continously as the coupling is varied
from the BCS to the BEC-limit. An important ingredient in these results is their
account for the repulsive interaction between strongly bound dimers in the BEC-limit
with scattering length $a_{dd}\approx 0.60\, a>0$ \cite{Petrov1}.
This interaction is missing in the early qualitative descriptions
of the $T=0$ BCS-BEC crossover problem by Eagles \cite{Eagles} and Leggett
\cite{Leggett}, which are based on using the standard BCS-groundstate
as a variational Ansatz for \textit{arbitrary} coupling \cite{Dukelsky}.
Beyond a purely numerical approach, the BCS-BEC crossover
problem has recently become amenable also to analytical methods via
an $\epsilon=4-d\,$ expansion \cite{Nishida31_07_06}.
It is based on the observation \cite{Nussinov:2005} that at the unitarity
point in $d=4$ (i.e.\ the point where a two-particle bound state appears)
the two-component Fermi gas is in fact an ideal Bose gas, because
a zero range interaction in $d=4$ can bind a state only at infinitely strong attraction.
In two dimensions, in turn, binding appears at arbitrary small couplings
and the unitary Fermi gas in $d\leq 2$ coincides with a non-interacting one
\cite{Nussinov:2005}. Within a field theoretic description, the physically interesting
3D problem can thus be approached by extrapolating expansions from the
upper and lower critical dimensions $d=4$ and $d=2$ respectively \cite{Nishida15_08_06}.
At finite temperature, numerical calculations are available for the
thermodynamics at the unitarity point. They are based on an auxiliary field
quantum Monte-Carlo method for the continuum problem \cite{Bulgac}
and on a diagrammatic determinant
Monte-Carlo method for the negative $U\/$-Hubbard model \cite{Burovski}.
Field-theoretic results at finite temperature, which open the possibility for
controlled and systematic expansions for the crossover thermodynamics
have been obtained very recently by Nishida \cite{Nishida15_08_06} within
an expansion around both the upper or lower critical dimension and by
Nikolic and Sachdev \cite{Nikolic} within a $1/N\,$ expansion
for a $2N\,$ component Fermi gas.

Our aim in the following is to present
a self-consistent many-body theory for the thermodynamics of resonantly
interacting fermions at arbitrary temperatures and detuning,
which directly addresses the physically relevant case of a
three dimensional, two-component Fermi gas.
The theory is based on a conserving, so-called
$\Phi$-derivable approach to the many-body problem, in which the
exact one- or two-particle Green functions serve as an infinite set of
variational parameters. It is an extension of
earlier work by one of  us \cite{Haussmann1,Haussmann2,Haussmann3}
and employs a combination of the
Luttinger-Ward and DeDominicis-Martin approach for obtaining the grand
canonical potential and the entropy, respectively. The condition of
gaplessness is enforced by a modified coupling constant, thus
accounting for the proper low energy behaviour in terms of a
Bogoliubov-Anderson mode.
We provide quantitative results for the critical temperature,
the equation of state and the entropy
near the  Feshbach-resonance as a function of both $T/T_F$ and
$1/k_Fa$. In spite of the fact that the critical behaviour at the
continuous superfluid transition is not captured correctly in our
approach, which gives rise to a weak first order transition,
the results provide a quantitative and consistent picture of the  
crossover which obeys thermodynamic relations at the
percent level. Our variational method is complementary both to
purely numerical and to field theoretic approaches to the problem.
The results can be used e.g.\ to predict the final temperature
reached after an adiabatic ramp across the Feshbach-resonance
starting deeply in the BEC-regime \cite{Carr} or to determine the size  
of the atom cloud in a harmonic trap near unitarity as a function of  
temperature.

The paper is organized as follows: in Sec.\ \ref{section_2} we introduce
our model and the basic many-body formalism
necessary for deriving a set of self-consistent equations for the
Green and vertex functions which are the variational
parameters of the theory. The complete
thermodynamics is then determined by integrals
of the momentum and frequency dependent Green functions.
It is shown that with a modified coupling constant,
the theory can be formulated in a way consistent
with Ward identities, which guarantees a gapless
Bogoliubov-Anderson mode  for arbitrary  strength of
the coupling. In Sec.\ \ref{sectionNR} we discuss the
numerical solution, providing quantitative results for
the critical temperature, the pressure, internal energy and the entropy
of the BCS-BEC crossover both in the normal and
superfluid phase. They are compared both with
experimental and theoretical results based on
numerical and field-theoretic approaches.
Finally in Sec.\ \ref{Discussion} we give a brief summary,
and indicate open problems.

\section{A many-body theory of resonantly interacting fermions}
\label{section_2}

In order to describe interacting fermions near a 
Fesh\-bach resonance, it is in general necessary to 
include the resonant, closed channel bound state
explicitely, e.g.\ within a Bose-Fermi-resonance model 
\cite{Holland2,Ohashi1}. As has been shown for instance by 
Diener and Ho \cite{Diener}, however, the situation can 
be simplified in the case of broad Feshbach resonances,
where the effective range $r^{\star}$ of the resonant interaction
is much smaller than both the background scattering 
length $a_{bg}$ and the Fermi wavelength $\lambda_F$. 
In this limit, which is in fact appropriate for the 
existing experimental studies of the BCS-BEC crossover
problem in $^{6}$Li \cite{Chin} and in $^{40}$K \cite{Greiner},
the problem can be reduced to a single channel Hamiltonian
with an instantaneous interaction \cite{Bruun,Diener,Simonucci,Burnett}. 
The associated effective two-body interaction is thus 
described by a pseudo-potential 
$V(\mathbf{r})\sim\delta(\mathbf{r})$ (appropriately
renormalized, see below) with a strength proportional to the 
scattering length
\begin{equation}
\label{eq:FBs-length}a=a_{bg}\left(1-\frac{\Delta B}{B-B_0}\right)\, .
\end{equation}
Here $a_{bg}$ is the off-resonant 
background scattering length  in the absence of the 
coupling to the closed channel while  $\Delta B$
and $B_0$ describe the width and position of the resonance
which may be tuned by an external magnetic field $B$. 
The interacting Fermi system is thus described by 
the standard Hamiltonian
\begin{equation}
\begin{split}
\hat H =& \int d^d r \sum_\sigma \frac{\hbar^2}{2m} [\nabla \psi^+_\sigma(\textbf{r})]
[\nabla \psi^{\ }_\sigma(\textbf{r})] \\ 
&+ \frac{1}{2} \int d^d r \int d^d r^\prime \sum_{\sigma\sigma^\prime} 
V(\textbf{r}-\textbf{r}^\prime) \\
&\hskip2.0cm \times \psi^+_\sigma(\textbf{r}) \psi^+_{\sigma^\prime}(\textbf{r}^\prime) 
\psi^{\ }_{\sigma^\prime}(\textbf{r}^\prime) \psi^{\ }_\sigma(\textbf{r}) \ , \\
\end{split}
\label{B_010}
\end{equation}
where $\psi^{\ }_\sigma(\textbf{r})$ and $\psi^+_\sigma(\textbf{r})$ are the usual fermion 
field operators. The formal spin index $\sigma$ labels two internal degrees of freedom, 
which in practice are two different hyperfine states. In the approximation, where the 
effective range of the resonant interaction is taken to zero, the interaction potential 
can formally be replaced by a delta potential between fermions of opposite spin
\begin{equation}
V(\textbf{r} - \textbf{r}^\prime) = g_0\, \delta(\textbf{r} - \textbf{r}^\prime)\, .
\label{B_020}
\end{equation}
Its strength $g_0$ needs to be renormalized for dimensions $d\geq 2$ by introducing the 
scattering amplitude $g$ via
\begin{equation}
\frac{1}{g} = \frac{1}{g_0} + \int \frac{d^dk}{(2\pi)^d} 
\ \frac{m}{\hbar^2 \textbf{k}^2} \ .
\label{B_030}
\end{equation}
For dimensions $d\geq 2$ the integral diverges at high momenta. Since the scattering amplitude
$g$ is kept constant, the bare interaction parameter $g_0$ must be taken
to zero in the limit where the cutoff diverges. The associated limiting 
process $g_0\rightarrow -0$ accounts for the replacement of the bare potential \eqref{B_020}
by a pseudopotential with the proper scattering length.  
While the formulas are derived for arbitrary space dimensions $d$, eventually we consider 
fermions for $d=3$. In this case the scattering amplitude $g$ is simply connected
to the s-wave scattering length $a$ given in \eqref{eq:FBs-length} by $g= 4\pi\hbar^2a/m$.

In the following, we consider a homogeneous situation 
described by a grand canonical distribution at fixed
temperature and chemical potential.   
The thermodynamic properties thus follow from
the grand partition function
\begin{equation}
Z = {\rm Tr}\{ \exp( -\beta [\hat H - \mu \hat N] ) \}
\label{B_050}
\end{equation}
and the associated grand  potential
\begin{equation}
\Omega = \Omega(T,\mu) = - \beta^{-1} \ln Z
\label{B_060}
\end{equation}
which is directly related to the pressure $p$ via $\Omega=-pV$.
Within our simplified model, 
where the range of the interaction is set to zero, the Fermi system is described by three
parameters: the temperature $T$, the chemical potential $\mu$ and the s-wave scattering
length $a$. Apart from an overall scale, the thermodynamics thus depends only on two 
dimensionless ratios. It is convenient to replace the chemical potential $\mu$ by the 
fermion density $n=k_F^3/3\pi^2$, which defines the Fermi wave number $k_F$ and the Fermi energy
$\varepsilon_F=\hbar^2k_F^2/2m$ as characteristic length and energy scales. The equilibrium state
is then uniquely determined by only two parameters: 
the dimensionless temperature $\theta = T / \varepsilon_F$ 
(we choose units for the temperature in which $k_B=1$), 
and the dimensionless interaction strength $v= 1/ k_F a$. In the special case
$B=B_0$ of an infinite scattering  length (the so-called unitarity limit),
the parameter $v$ drops out and the resulting thermodynamic quantities  
are universal functions of $\theta$ \cite{Ho}.

\subsection{Luttinger-Ward formalism}
\label{subsection_2A}

The BCS-BEC crossover is controlled by two physical phenomena.  The first one is 
connected with the formation of pairs due to the attractive interaction. The second one 
is the transition to superfluidity below a certain critical temperature $T_c$.
In the BCS-limit, the formation of pairs and the superfluid transition are simultaneous.
The transition is driven by the thermal breakup of pairs, i.e.\ by excitations
which may be described by a purely fermionic theory. With increasing strength of the 
interaction, however, there is an increasingly wide range of temperatures
where bound pairs coexist with unpaired fermions. In the BEC-limit, pair formation,
as a chemical equilibrium between bound and dissociated atoms, occurs 
at a temperature scale much higher than the superfluid transition. The 
latter is driven by collective excitations of a then purely bosonic system. A proper description 
of the crossover thus requires to account for both bosonic and fermionic excitations 
simultaneously.

Following the formalism developed by Luttinger and Ward \cite{LW60} for 
non-superfluid interacting Fermi systems, the grand thermodynamic potential \eqref{B_060} can be 
expressed as a unique functional of the Green function 
\begin{equation}
\begin{split}
G_{\sigma\sigma^\prime}&( \textbf{r} - \textbf{r}^\prime, \tau - \tau^\prime ) \\
=&
\begin{pmatrix}
\delta_{\sigma\sigma^\prime} {\cal G}( \textbf{r} - \textbf{r}^\prime, \tau - \tau^\prime ) 
&\varepsilon_{\sigma\sigma^\prime} {\cal F}( \textbf{r} - \textbf{r}^\prime, \tau - \tau^\prime ) \cr
-\varepsilon_{\sigma\sigma^\prime} {\cal F}^*( \textbf{r}^\prime - \textbf{r}, \tau - \tau^\prime ) 
&-\delta_{\sigma\sigma^\prime} {\cal G}( \textbf{r}^\prime - \textbf{r}, \tau^\prime - \tau ) \cr
\end{pmatrix}
\end{split}
\label{B_100}
\end{equation}
in the form
\begin{equation}
\Omega[G] = \beta^{-1} \bigl( - {\textstyle \frac{1}{2}} 
\text{Tr} \{ -\ln G + [G_0^{-1} G - 1] \} - \Phi[G] \bigr ) \ .
\label{B_110}
\end{equation}
The trace $\text{Tr}$ is defined with 
respect to the formal index $X=(\textbf{r}, \tau, \sigma, \alpha)$ which combines the space 
variable $\textbf{r}$, the imaginary time $\tau$, the spin index $\sigma$, and the Nambu index 
$\alpha$. The interaction between the fermions is described by the functional $\Phi[G]$, which 
can be expressed in terms of a perturbation series of irreducible Feynman-Diagrams where the 
propagator lines are dressed and identified by the matrix Green function $G$ of \eqref{B_100}.

While the formalism of Luttinger and Ward was originally derived for normal quantum liquids,
it is well suited also to describe superfluid systems. Indeed the nondiagonal elements of the 
matrix Green function $G$ represent the order parameter of the superfluid transition. The 
minimization of the grand potential $\Omega[G]$ as a functional of the Green function $G$ thus 
incorporates the standard thermodynamic criterion that the order parameter is found by minimizing 
the thermodynamic potential.  The stationarity condition 
\begin{equation}
\delta \Omega[G] / \delta G = 0  
\label{B_120}
\end{equation}
uniquely determines the full
matrix Green function $G$ of the interacting system and hence the order parameter.
It is important to note, that the thermodynamic potential $\Omega[G]$ depends
on the exact Green function $G$.  The formalism of Luttinger and Ward thus leads 
via \eqref{B_120} to a self-consistent theory for the matrix Green function $G$.
Since the Green functions contain information about the full dynamical 
behaviour via the imaginary time dependence of the Matsubara formalism,
the Luttinger-Ward approach not only provides results for the equilibrium 
thermodynamic quantities but also determines spectral functions and transport
properties. In our present work, however, dynamical properties will not 
be discussed. 

\begin{figure}
\includegraphics[width=0.45\textwidth]{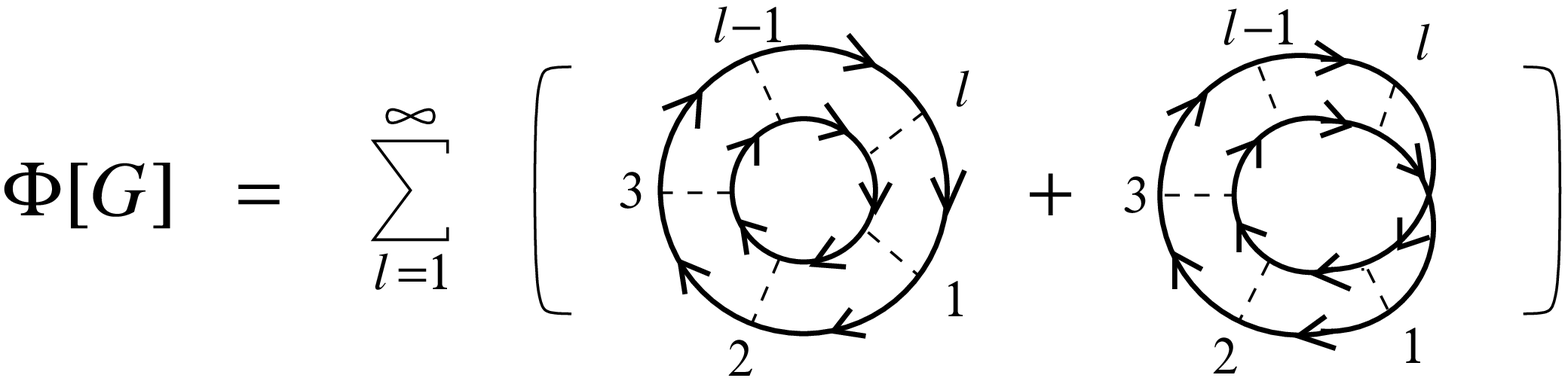}
\caption{The functional $\Phi[G]$ in self-consistent ladder approximation. The propagator lines
are dressed lines identified by the matrix Green function $G$.}
\label{FigLadderApprox}
\end{figure}

The functional $\Phi[G]$ is defined by an infinite perturbation series of irreducible Feynman
diagrams and an exact expression for $\Phi[G]$ is clearly beyond what can be done analytically. 
An approximation which properly describes the formation of pairs, is a ladder 
approximation \cite{FW71}. In Fig.\ \ref{FigLadderApprox}, the related diagrams 
of $\Phi[G]$ are shown. The ladder approximation is self consistent because the propagator lines 
are dressed lines which are identified by the matrix Green function $G$. In the 
weak coupling BCS regime the ladder approximation becomes exact. For very strong attractive 
interactions, well above the pairing threshold, the fermion system is a Bose liquid of dilute atom 
pairs. In this limit the ladder approximation describes the formation of pairs (two-particle
problem) exactly, however the interaction between the pairs (four-particle problem) only 
approximately \cite{Petrov1,Pieri00}. In particular the resulting dimer-dimer scattering
length is given by the Born approximation $a_{dd}^{(B)}=2a$.

\subsection{DeDominicis-Martin formalism}
\label{subsection_2B}

An extension of the Luttinger-Ward formalism was given by DeDominicis and Martin \cite{DM64}.
They introduce up to four external fields, which couple to products of one, 
two, three, and four field operators, and perform the Legendre transformations to the 
corresponding conjugate variables - the Green functions. For
fermion systems only two external fields are relevant which couple to even products of
fermion field operators. The related two conjugate variables of the Legendre transformation 
are the one-particle Green function $G$ and the two-particle Green function $G_2$. 
Within our approach below, the second Legendre transformation is performed explicitely. 
A more convenient conjugate variable is then the vertex function $\Gamma$ which is related
to $G_2$ by \eqref{B_127} below. 
Thus, DeDominicis and Martin obtain a thermodynamic potential which is a functional of both 
$G$ and $\Gamma$. More precisely, it turns out that the relevant functional is the entropy 
$S = F^{(2)}$ where
\begin{equation}
\begin{split}
F^{(2)}[G,\Gamma] = &{\textstyle \frac{1}{2}} \text{Tr} 
\{ -\ln G + [(-i\hbar\omega_n) G - 1] \} \\
&+ {\textstyle \frac{1}{2}} \text{Tr} \{ \ln [ 1 - {\textstyle \frac{1}{2}} \bar \Gamma ] 
+ {\textstyle \frac{1}{2}} \bar \Gamma 
+ {\textstyle \frac{1}{2}} [ {\textstyle \frac{1}{2}} \bar \Gamma ]^2 \\
&- (1/4!) [ \bar \Gamma ]^2 \} 
+ {\cal K}^{(2)}[G,\Gamma] 
\end{split}
\label{B_123}
\end{equation}
(see (61) in the second paper of Ref.\ \onlinecite{DM64} and identify $G_1=G$, $C_2=-\Gamma$,
and $\bar C_2= -\bar \Gamma$ therein). 
$\bar \Gamma$ is defined in \eqref{B_128} below. 

The formalism of DeDominicis and Martin is ideally adapted to describe the BCS-BEC crossover 
because it explicitly deals with the one-particle Green function $G$, which represent
the properties of the single fermions, and the vertex function $\Gamma$, 
which describes the eventually purely bosonic properties 
of the fermion pairs (both condensed or noncondensed). In particular, a full 
implementation of their formalism is needed to correctly account for four particle 
correlation, which is necessary to obtain the exact result $a_{dd}=0.60 \, a$ for 
the dimer-dimer scattering length in the BEC limit.

As in standard thermodynamics, the
entropy \eqref{B_123} is maximized under the constraints that all conserved quantities are 
kept constant. For the interacting fermion system defined by the Hamiltonian \eqref{B_010}
the conserved quantities are the internal energy $U=\langle \hat H \rangle$ and the particle
number $N=-{\textstyle \frac{1}{2}} \text{Tr} \{ G \}$. Evaluating the thermal average of the
Hamiltonian \eqref{B_010} we find that $U$ can be expressed in terms of $G$ and $\Gamma$ 
(see \eqref{B_127} and \eqref{B_200} below).  

Consequently, the entropy $F^{(2)}[G,\Gamma]=S[G,\Gamma]$, the internal energy $U[G,\Gamma]$, 
and the particle number $N[G]$ are functionals depending on $G$ and $\Gamma$. In order to find 
the maximum of the entropy under the constraint of given average values of the particle number 
and the internal energy, DeDominicis and Martin \cite{DM64} consider the functional 
\begin{equation}
W[G,\Gamma] = F^{(2)}[G,\Gamma] - \lambda_U U[G,\Gamma] - \lambda_N N[G]
\label{B_124}
\end{equation}
where $\lambda_U$ and $\lambda_N$ are two Lagrange parameters for the two constraints.
Alternatively and equivalently, we consider the functional
\begin{equation}
\Omega[G,\Gamma] = U[G,\Gamma] - T\, S[G,\Gamma] - \mu\, N[G]
\label{B_125}
\end{equation}
which is the grand thermodynamic potential where the temperature $T$ and the chemical potential
$\mu$ are the Lagrange parameters. Both functionals \eqref{B_124} and \eqref{B_125} must be 
stationary under small variations of $G$ and $\Gamma$. In this way, we obtain the stationarity
criteria
\begin{equation}
\delta \Omega[G,\Gamma] / \delta G = 0 \enskip \mbox{and} 
\enskip \delta\Omega[G,\Gamma] / \delta \Gamma = 0
\label{B_126}
\end{equation}
which uniquely determine the one-particle Green function $G$ and the vertex function $\Gamma$.

\begin{figure}[b]
\includegraphics[width=0.45\textwidth]{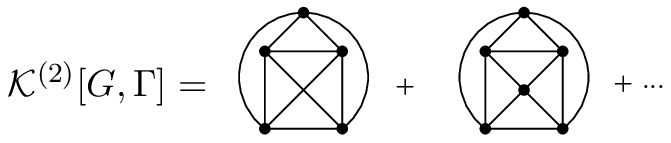}
\caption{The functional ${\cal K}^{(2)}[G,\Gamma]$ is the sum of all 2-line 
irreducible diagrams. The propagator lines and the vertices (full circles) are dressed 
and identified with $G$ and $\Gamma$, respectively.}
\label{FigK_2}
\end{figure}

In order to simplify the second trace in the entropy functional \eqref{B_123} 
it is convenient to define a modified vertex function $\bar \Gamma$ by 
\begin{equation}
\bar \Gamma_{X_1 X_2 X_3 X_4} = G^{1/2}_{X_1 Y_1} G^{1/2}_{X_2 Y_2} \Gamma_{Y_1 Y_2 Y_3 Y_4} 
G^{1/2}_{Y_3 X_3} G^{1/2}_{Y_4 X_4} 
\label{B_128}
\end{equation}
where the four external propagator lines are amputated only half way (see (46) in the second
paper of \onlinecite{DM64}). For a proper definition of the second trace and the related 
matrix products the four indices of the modified vertex function 
must be grouped into pairs according to $\bar \Gamma = \bar \Gamma_{(X_1 X_2) (X_3 X_4)}$.
The last term in \eqref{B_123}, the functional ${\cal K}^{(2)}[G,\Gamma]$ (depicted in 
Fig.\ \ref{FigK_2}), is defined by an infinite perturbation series of 2-line irreducible 
Feynman diagrams, where the propagator lines and the vertices are dressed and identified 
by the one-particle Green function $G$ and by the vertex function $\Gamma$, respectively.

In order to understand the physical meaning of the various contributions to the
thermodynamic potential, we note that the Luttinger-Ward formalism and the 
DeDominicis-Martin formalism are related to each other by a Legendre transformation, 
in which the bare two-particle interaction as an external field is transformed into 
the two-particle Green function $G_2$. This Legendre transformation may
be interpreted as a renormalization procedure. Since the two-particle Green function $G_2$ 
is expressed in terms of the vertex function $\Gamma$ by 
\begin{equation}
\begin{split}
G_{2,X_1 X_2 X_3 X_4} = & G_{X_1 X_3} G_{X_2 X_4} \\
&- G_{X_1 X_4} G_{X_2 X_3} - G_{X_1 X_2} G_{X_3 X_4} \\
&- G_{X_1 Y_1} G_{X_2 Y_2} \Gamma_{Y_1 Y_2 Y_3 Y_4} G_{Y_3 X_3} G_{Y_4 X_4} \ ,
\end{split}
\label{B_127}
\end{equation}
the bare two-particle interaction is replaced by the vertex function $\Gamma$ as the 
renormalized interaction, which is the many-particle generalization of the scattering 
amplitude $g$. 

We may compare the functionals \eqref{B_110} and \eqref{B_123} directly with each other. The first 
trace in \eqref{B_123} is identified by the trace in \eqref{B_110}, which describe the 
contribution of single particles to the grand canonical potential and entropy, respectively.
The second trace and the functional ${\cal K}^{(2)}[G,\Gamma]$ in 
\eqref{B_123} which represent the interaction terms are related to the functional $\Phi[G]$ in 
\eqref{B_110}. By close inspection (see (62) in Ref.\ \onlinecite{DM64}) we find that the second 
trace in \eqref{B_123} represents the inverted perturbation series of ladder diagrams. It includes
both particle-particle and particle-hole ladders,  which describe the scattering and formation of 
pairs and it also includes bubble diagrams, which describe the screening of the interaction.

These three types of diagrams and mixtures of them arise because the vertices are 
symmetrized so that each of them can be expressed as a sum of three unsymmetrized vertices.
As a result,  the self-consistent ladder approximation of the functional $\Phi[G]$ shown in 
Fig.\ \ref{FigLadderApprox} can be reformulated within the formalism of DeDominicis and Martin
in the following way: the second trace is approximated by keeping only the particle-particle
ladders and the complicated functional ${\cal K}^{(2)}[G,\Gamma]$ is set equal to zero. This 
neglects the screening of the interaction due to particle-hole excitations (see 
Subsec.\ \ref{subsection_ct} below) and also the coupling  between collective excitations 
and bound pairs.

In the following subsections we employ the formalism of DeDominicis and Martin to construct 
explicit expressions for $S[G,\Gamma]$, $U[G,\Gamma]$, and $N[G]$. From \eqref{B_125} we obtain
the functional $\Omega[G,\Gamma]$. The stationarity criteria \eqref{B_126} imply two self-consistent
equations for the Green function $G$ and the vertex function $\Gamma$. Solving the second equation
with respect to $\Gamma$ and inserting the resulting vertex function into $\Omega[G,\Gamma]$ 
we recover the functional $\Omega[G]$ of the Luttinger-Ward formalism together with the stationarity 
condition \eqref{B_120}. This fact explicitely demonstrates the equivalence of the Luttinger-Ward 
and DeDominics-Martin formalism for our approximation scheme (see \eqref{B_430}-\eqref{B_450} below) 
once the appropriate stationarity conditions have been taken into account.

\begin{widetext}
\subsection{Thermodynamic potentials}
\label{subsection_2C}
The formalism of Luttinger and Ward \cite{LW60} allows to calculate directly the grand 
thermodynamic potential $\Omega$. The functional $\Phi[G]$ has been evaluated explicitly 
in Ref.\ \onlinecite{Haussmann3}. Inserting this result into \eqref{B_110} we obtain
\begin{equation}
\begin{split}
\Omega[G] = &-L^d \int \frac{d^dk}{(2\pi)^d} \ \frac{1}{\beta}
\sum_{\omega_n} \text{Tr} \bigl\{ -\ln[ G(\textbf{k},\omega_n) ] +
[ G_0(\textbf{k},\omega_n)^{-1} G(\textbf{k},\omega_n) - 1 ] \bigr\} \\
&+ L^d\, g_0 \vert {\cal F}(\textbf{0},0) \vert^2 
\ +\ \frac{1}{2}\, L^d \int \frac{d^dK}{(2\pi)^d} \ \frac{1}{\beta} \sum_{\Omega_n}
\text{Tr} \bigl\{ \ln[ 1 + g_0\, \chi(\textbf{K},\Omega_n) ] \bigr\} \ . \\
\end{split}
\label{B_130}
\end{equation}
In this formula the matrix Green function is defined by
\begin{equation}
G(\textbf{k},\omega_n) = \bigl( G_{\alpha \alpha^\prime}(\textbf{k},\omega_n) \bigr) = 
\begin{pmatrix}
{\cal G}(\textbf{k},\omega_n) &{\cal F}(\textbf{k},\omega_n)\cr
{\cal F}(\textbf{k},\omega_n)^* &-{\cal G}(\textbf{k},\omega_n)^* 
\end{pmatrix}
\ .
\label{B_140}
\end{equation}
Knowledge of the matrix Green functions determines the matrix pair propagator via
\begin{equation}
\chi(\textbf{K},\Omega_n) = \bigl( \chi_{\alpha \alpha^\prime}(\textbf{K},\Omega_n)
\bigr) = \Bigl( \int \frac{d^dk}{(2\pi)^d} \ \frac{1}{\beta} \sum_{\omega_n}
G_{\alpha \alpha^\prime}(\textbf{K}-\textbf{k},\Omega_n-\omega_n)
G_{\alpha \alpha^\prime}(\textbf{k},\omega_n) \Bigr) \ .
\label{B_150}
\end{equation}
\end{widetext}
In order to distinguish between fermionic and bosonic functions the fermionic
wave vectors and Matsubara frequencies are denoted by small letters, while the bosonic
wave vectors and Matsubara frequencies are denoted by capital letters. In the second term
of \eqref{B_130} the anomalous Green function is identified by ${\cal F}(\textbf{0},0)=
{\cal F}(\textbf{r}=\textbf{0},\tau=0)$.
The formulas are derived for an arbitrary dimension of space $d$. The volume is assumed
to be a cube with edge length $L$ and periodic boundary conditions, where the limit 
$L\rightarrow \infty$ is taken.

The strength of the attractive interaction is included by the bare interaction parameter
$g_0$. The kinetic energy of the atoms $\varepsilon_\textbf{k}= \hbar^2 \textbf{k}^2 /2m$ and 
the chemical potential $\mu$ are implicitly included via the free matrix Green function
$G_0(\textbf{k},\omega_n)$, which is related to the free normal Green function 
\begin{equation}
{\cal G}_0(\textbf{k},\omega_n) = 1 / [-i \hbar\omega_n + \varepsilon_\textbf{k} - \mu ]
\label{B_160}
\end{equation}
and the free anomalous Green function
\begin{equation}
{\cal F}_0(\textbf{k},\omega_n) = 0
\label{B_170}
\end{equation}
by a formula which is analogous to \eqref{B_140}. The temperature $T$ is included explicitly
by the factors $1/\beta$ and implicitly by the Matsubara frequencies $\omega_n$ and $\Omega_n$.

As evident from \eqref{B_130}, the formalism of Luttinger and Ward, though 
including the exact single-particle Green function, still contains the bare coupling constant 
$g_0$. In the DeDominicis-Martin formalism the bare coupling is renormalized
and replaced by the exact vertex function $\Gamma$ via a second Legendre transformation.
The corresponding functional $ S[G,\Gamma] = F^2[G,\Gamma]$, is
just the dimensionless entropy as given in (\eqref{B_123}) 

As discussed above we restrict the second trace in \eqref{B_123} to the particle-particle 
ladders and by the nature of our interaction potential \eqref{B_020} to s-wave scattering. 
Furthermore, we omit the 2-line irreducible Feynman diagrams by setting ${\cal K}^{(2)}[G,\Gamma] = 0$.
This approximation covers the essential features of the crossover problem namely the formation
of pairs and their condensation. Within our ladder approximation, the DeDominicis-Martin formalism 
thus leads to an expression for the entropy of the form
\begin{widetext}
\begin{equation}
\begin{split}
S[G,\Gamma] =& \beta\, L^d \int \frac{d^dk}{(2\pi)^d} \ \frac{1}{\beta}
\sum_{\omega_n} \text{Tr} \bigl\{ -\ln[ G(\textbf{k},\omega_n) ] +
[ -i\hbar\omega_n G(\textbf{k},\omega_n) - 1 ] \bigr\} \\
&+ \frac{1}{2}\, \beta\, L^d \int \frac{d^dK}{(2\pi)^d} \ \frac{1}{\beta}
\sum_{\Omega_n} \text{Tr} \bigl\{ \ln[ 1 - \chi(\textbf{K},\Omega_n) \Gamma(\textbf{K},\Omega_n) ]
+ \chi(\textbf{K},\Omega_n) \Gamma(\textbf{K},\Omega_n) \bigr\} \ . \\
\end{split}
\label{B_174}
\end{equation}
\end{widetext}
The first term is clear. It is directly obtained from the first trace in \eqref{B_123}.
However, the second term resulting from the second trace in \eqref{B_123}
needs further explanation. From \eqref{B_128} and the definition of the pair propagator 
\eqref{B_150} we infer
\begin{equation}
\begin{split}
\bar \Gamma(\textbf{K},\Omega_n) 
&= \chi(\textbf{K},\Omega_n)^{1/2} \, \Gamma(\textbf{K},\Omega_n) \, 
\chi(\textbf{K},\Omega_n)^{1/2} \\
&= \chi(\textbf{K},\Omega_n) \, \Gamma(\textbf{K},\Omega_n) \\
&= \Gamma(\textbf{K},\Omega_n) \, \chi(\textbf{K},\Omega_n) \ .
\end{split}
\label{B_176}
\end{equation}
The reduction to particle-particle ladders implies that the 
Nambu indices are pairwise identical. In this way, the four Nambu indices of the vertex function 
$\Gamma$ reduce to two Nambu indices. As a result, the vertex function $\Gamma(\textbf{K},\Omega_n)=
(\Gamma_{\alpha\alpha^\prime}(\textbf{K},\Omega_n))$ is a $2\times 2$ matrix in the Nambu space
similar to the matrix Green function \eqref{B_140}. For the formalism of Luttinger and Ward 
the reduction of the vertex is described in detail in Ref.\ \onlinecite{Haussmann3} and also 
in Ref.\ \onlinecite{Haussmann1}.
Since the second trace of \eqref{B_123} is reduced to the particle-particle ladders and the
structure of the vertex function is simplified considerably due to s-wave scattering, the prefactors 
of the terms in the second trace of \eqref{B_174} are changed. The factor $\frac{1}{2}$ in front of 
$\bar \Gamma$ disappears. Furthermore, the quadratic terms in the second trace cancel.

Another important quantity to consider is the internal energy $U$. With the help of the 
delta potential \eqref{B_020} we find for the expectation value of the Hamiltonian \eqref{B_010},
\begin{equation}
\begin{split}
U =& \int d^d r \sum_\sigma \frac{\hbar^2}{2m} \, \langle [\nabla \psi^+_\sigma(\textbf{r})]
[\nabla \psi^{\ }_\sigma(\textbf{r})] \rangle \\ 
&+ \frac{1}{2} \int d^d r  \sum_{\sigma \sigma^\prime} g_0 \,
\langle \psi^+_\sigma(\textbf{r}) \psi^+_{\sigma^\prime}(\textbf{r}) 
\psi^{\ }_{\sigma^\prime}(\textbf{r}) \psi^{\ }_\sigma(\textbf{r}) \rangle \ . \\
\end{split}
\label{B_190}
\end{equation}
The second term contains an average of four fermion field operators 
which can be expressed in terms of the two-particle Green function $G_2$. Following the 
formalism of DeDominicis and Martin \cite{DM64} and using \eqref{B_127} the two-particle Green 
function can be expressed by four terms. The first three terms represent the three possibilities 
to factorize the two-particle Green function into products of two one-particle Green functions 
according to the Wick theorem. These terms provide the Hartree energy, the Fock energy, and 
the Bogoliubov energy. The fourth term is the connected part of the two-particle Green function 
and provides the correlation energy. Taking all terms together we obtain the internal energy
\begin{widetext}
\begin{equation}
\begin{split}
U[G,\Gamma] = &-2\, L^d \int \frac{d^dk}{(2\pi)^d} \ \varepsilon_\textbf{k} 
\ {\cal G}(\textbf{k},\tau=-0) \ +\ L^d\, g_0 \vert {\cal F}(\textbf{0},0) \vert^2 \\
&+ \frac{1}{2}\, L^d \int \frac{d^dK}{(2\pi)^d} \ \frac{1}{\beta} \sum_{\Omega_n}
g_0\, \text{Tr} \bigl\{ \chi(\textbf{K},\Omega_n) 
- \chi(\textbf{K},\Omega_n)\, \Gamma(\textbf{K},\Omega_n)\, \chi(\textbf{K},\Omega_n) 
\bigr\} \ . \\
\end{split}
\label{B_200}
\end{equation}
\end{widetext}

Finally, the particle number $N$ is defined by the average
\begin{equation}
N = \langle \hat N \rangle = 
\int d^d r \sum_\sigma \langle \psi^+_\sigma(\textbf{r}) \psi^{\ }_\sigma(\textbf{r}) \rangle 
\label{B_210}
\end{equation}
which can be expressed in terms of the normal Green function in the standard form
\begin{equation}
N[G] = -2\, L^d \int \frac{d^dk}{(2\pi)^d} \ {\cal G}(\textbf{k},\tau=-0) \ .
\label{B_220}
\end{equation}

The entropy \eqref{B_174}, the internal energy \eqref{B_200}, and the particle number \eqref{B_220}
are the basic functionals of the formalism of DeDominicis and Martin. The entropy 
$S[G,\Gamma]$ is maximized under the constraints that the internal energy $U[G,\Gamma]$ and
the particle number $N[G]$ are constant. In order to do this, the grand thermodynamic potential
$\Omega[G,\Gamma]$ is defined by \eqref{B_125} where the temperature $T$ and the chemical potential
$\mu$ are Lagrange parameters. The self-consistent equations for the Green function $G$ and the
vertex function $\Gamma$ are obtained from the stationarity conditions \eqref{B_126}. Formally,
the formalism of DeDominicis and Martin yields a different expression for the grand thermodynamic 
potential $\Omega$ than the formalism of Luttinger and Ward does by \eqref{B_130}. However, 
it can be shown that the results are identical if and only if $G$ and $\Gamma$ satisfy the 
self-consistent equations (see end of Subsec.\ \ref{subsection_2E}).

The functionals \eqref{B_174}, \eqref{B_200}, and \eqref{B_220} do not depend explicitly on the 
thermodynamic parameters $T$ and $\mu$. While the temperature appears explicitely via the
factor $\beta = 1/T$ and the Matsubara frequency $\omega_n\sim \Omega_n\sim T$, a proper 
rescaling of the functions $G\rightarrow \beta\,G$, $\chi\rightarrow \beta\,\chi$, and 
$\Gamma\rightarrow \Gamma$ implies that all factors $\beta$ and $T$ cancel in all three 
functionals. The temperature $T$ and the chemical potential $\mu$ enter only as Lagrange 
parameters via the constraints. This fact is a general property of the formalism of DeDominicis 
and Martin. The fermion mass $m$, the kinetic energy $\varepsilon_\textbf{k}=
\hbar^2 \textbf{k}^2 /2m$, and the interaction parameter $g_0$, which determine the microscopic 
properties of the interacting fermion system, are present only in the internal energy functional 
\eqref{B_200}.

An alternative expression for the entropy is obtained from the grand thermodynamic potential 
of Luttinger and Ward \eqref{B_130} according to the standard thermodynamic relation
\begin{equation}
S = - \partial \Omega / \partial T \ .
\label{B_260}
\end{equation}
Taking the partial derivative we obtain an expressions which formally differs from \eqref{B_174}. 
However, provided that $G$ and $\Gamma$ satisfy the self-consistent equations, the results for 
the entropy will be identical. Therefore, both the Luttinger-Ward
and the DeDominicis-Martin formalism exactly obey all the standard
thermodynamic relations provided the Green functions obey the
stationarity conditions \eqref{B_120} and \eqref{B_126}.
The equivalence of the different formal expressions in thermal equilibrium is very important for 
the consistency of our theory and the compatibility of the self-consistent ladder approximation 
for all thermodynamic quantities. Apart from the entropy, we can also determine the pressure 
$p=-\Omega/L^d$ as a functional of the Green function $G$ using \eqref{B_130} or \eqref{B_125}.
The dimensionless thermodynamic quantities $\Omega/N\varepsilon_F$, $U/N\varepsilon_F$ and $S/N$ 
will be calculated numerically in Sec.\ \ref{sectionNR} and discussed in the following sections.

\subsection{Self-consistent equations for the Green and vertex functions}
\label{subsection_2D}

The self-consistent equations for the Green functions follow 
directly from  the stationarity condition \eqref{B_120}. 
Inserting the general functional of the Luttinger-Ward formalism \eqref{B_110} into this
condition we obtain the Dyson equation 
\begin{equation}
G^{-1}_{\alpha \alpha^\prime}(\textbf{k},\omega_n) = 
G^{-1}_{0,\alpha \alpha^\prime}(\textbf{k},\omega_n) 
- \Sigma_{\alpha \alpha^\prime}(\textbf{k},\omega_n) \ .
\label{B_290}
\end{equation}
The self energy $\Sigma$ is identified by the functional derivative
\begin{equation}
\Sigma_{\alpha \alpha^\prime}(\textbf{k},\omega_n) = - \, \frac{1}{\beta L^d} \,
\frac{\delta \Phi[G]}{\delta G_{\alpha^\prime \alpha}(\textbf{k},\omega_n)} \ .
\label{B_300}
\end{equation}
The functional $\Phi[G]$ is defined by a perturbation series. The related Feynman diagrams are 
shown in Fig.\ \ref{FigLadderApprox} for the self-consistent ladder approximation. Inserting
the grand thermodynamic potential \eqref{B_130} into the constraint \eqref{B_120}, we obtain
an explicit expression for the self energy which is
\begin{equation}
\begin{split}
\Sigma_{\alpha \alpha^\prime}(\textbf{r},\tau) = 
&\Sigma_{1,\alpha \alpha^\prime} \, \delta(\textbf{r}) \, \delta_F(\tau/\hbar) \\
&+ G_{\alpha^\prime \alpha}(-\textbf{r},-\tau) 
\Gamma_{\alpha \alpha^\prime}(\textbf{r},\tau) \ . \\
\end{split}
\label{B_310}
\end{equation}
In the first term $\delta_F(\tau/\hbar)$ is the fermionic delta function 
which is antiperiodic. The order parameter of the superfluid transition 
\begin{equation}
\Delta = g_0\, {\cal F}(\textbf{0},0)
\label{B_320}
\end{equation}
is represented by the nondiagonal elements of the matrix
\begin{equation}
\Sigma_1 =
\begin{pmatrix}
0 &\Delta \\
\Delta^* &0 \\
\end{pmatrix} \ .
\label{B_330}
\end{equation}
In the second term of \eqref{B_310} $\Gamma$ is the matrix vertex function, which is related
to the matrix pair propagator $\chi$ by 
\begin{equation}
\Gamma^{-1}_{\alpha \alpha^\prime}(\textbf{K},\Omega_n) = g_0^{-1} \delta_{\alpha \alpha^\prime} 
+ \chi_{\alpha \alpha^\prime}(\textbf{K},\Omega_n) \ .
\label{B_335}
\end{equation}
Eventually, $\chi$ is represented in terms of the matrix Green function by \eqref{B_150}. 
Eq.\ \eqref{B_335} is just the Bethe-Salpeter equation in ladder approximation. It is 
responsible for the fact that the binding of fermion pairs is described appropriately. 
Taken together, we have now a set of self-consistent equations for the matrix Green function 
$G$ and the matrix vertex function $\Gamma$ which have to be solved numerically.

Alternatively we can derive the self-consistent equations by inserting the functional \eqref{B_125}
of the formalism of DeDominicis and Martin into the related stationarity conditions \eqref{B_126}.
We obtain the Dyson equation \eqref{B_290} from the first condition and the Bethe-Salpeter equation
\eqref{B_335} from the second condition. In this way we prove that both the Luttinger-Ward formalism
and the DeDominicis-Martin formalism are equivalent within our approximation.

Unfortunately, in the present form, the matrix pair propagator $\chi$ defined 
in \eqref{B_150} is divergent. While the 
sum over the Matsubara frequencies is finite, the integral over the wave vector is ultraviolet
divergent for dimensions $d\geq 2$. For this reason a renormalization is necessary. We define
the regularized pair propagator by
\begin{widetext}
\begin{equation}
M_{\alpha \alpha^\prime}(\textbf{K},\Omega_n) = \int \frac{d^dk}{(2\pi)^d} 
\Bigl[ \frac{1}{\beta} \sum_{\omega_n}
G_{\alpha \alpha^\prime}(\textbf{K}-\textbf{k},\Omega_n-\omega_n)
G_{\alpha \alpha^\prime}(\textbf{k},\omega_n) - \frac{m}{\hbar^2 \textbf{k}^2} 
\, \delta_{\alpha \alpha^\prime} \Bigr] \ .
\label{B_340}
\end{equation}
\end{widetext}
Inserting this formula into \eqref{B_335} we obtain the renormalized Bethe-Salpeter equation
\begin{equation}
\Gamma^{-1}_{\alpha \alpha^\prime}(\textbf{K},\Omega_n) = g^{-1} \delta_{\alpha \alpha^\prime}
+ M^{\ }_{\alpha \alpha^\prime}(\textbf{K},\Omega_n) \ .
\label{B_350}
\end{equation}
The bare interaction strength $g_0$ is renormalized according to \eqref{B_030} and replaced 
by the scattering amplitude $g$. For $d=3$ dimensions $g$ is expressed in terms of the
s-wave scattering length $a$ by $g = 4\pi\hbar^2a/m$. 

The zero range of the interaction between the fermions implies that 
\begin{equation}
{\cal F}(\textbf{0},0) = \int \frac{d^dk}{(2\pi)^d} \ \frac{1}{\beta} \sum_{\omega_n}
{\cal F}(\textbf{k},\omega_n)
\label{B_380}
\end{equation}
is infinite. For this reason the order-parameter formula \eqref{B_320} must be renormalized,
too. Replacing the bare interaction strength $g_0$ by the scattering amplitude $g$ 
according to \eqref{B_030}, we obtain the renormalized formula
\begin{equation}
\Delta = g \int \frac{d^dk}{(2\pi)^d} \Bigl[ {\cal F}(\textbf{k},\tau=0) +
\Delta\, \frac{m}{\hbar^2 \textbf{k}^2} \Bigr] \ .
\label{B_390}
\end{equation}
Here, the integral over the wave vector is finite.

\subsection{Reformulation in terms of mean-field Green functions}
\label{subsection_2E}

In mean-field approximation the self energy $\Sigma(\textbf{k},\omega_n)$ is replaced by 
$\Sigma_1$ defined in \eqref{B_330}. Since $\Sigma_1$ depends neither on wavevector 
nor on frequency, the approximation $\Sigma \approx \Sigma_1$ just describes the 
formation of a pair condensate within a BCS-type mean-field theory where the destruction of
superfluidity is driven by the breakup of pairs. This is the correct description in the 
weak coupling limit, however for strong coupling
the superfluid transition is driven by finite momentum pairs 
whose contribution is contained in the second term of the self energy \eqref{B_310}.
Inserting the mean-field self energy into the Dyson equation \eqref{B_290} we obtain
\begin{equation}
\begin{split}
G_1(\textbf{k},\omega_n)^{-1} =& G_0(\textbf{k},\omega_n)^{-1} - \Sigma_1 \\
=&
\begin{pmatrix}
-i \omega_n + (\varepsilon_\textbf{k} - \mu) &-\Delta \\
-\Delta^* &-i \omega_n - (\varepsilon_\textbf{k} - \mu) \\
\end{pmatrix}
\end{split}
\label{B_400}
\end{equation}
where $G_1$ is the matrix Green function in mean-field approximation. 

If we consider the self-consistent equations and the formulas for the thermodynamic potentials
we realize that the spectrum $\varepsilon_\textbf{k}$ of the fermionic atoms and 
the chemical potential $\mu$ enter the formulas only implicitly via the free matrix Green 
function $G_0$. We can transform the formulas so that $G_0$ is replaced in favor of the
mean-field matrix Green function $G_1$. As a result we obtain the Dyson equation
\begin{equation}
G^{-1}_{\alpha \alpha^\prime}(\textbf{k},\omega_n) = 
G^{-1}_{1,\alpha \alpha^\prime}(\textbf{k},\omega_n) 
- \tilde \Sigma_{\alpha \alpha^\prime}(\textbf{k},\omega_n)
\label{B_410}
\end{equation}
where
\begin{equation}
\tilde \Sigma_{\alpha \alpha^\prime}(\textbf{r},\tau) = 
G_{\alpha^\prime \alpha}(-\textbf{r},-\tau) \Gamma_{\alpha \alpha^\prime}(\textbf{r},\tau)
\label{B_420}
\end{equation}
is the second term of the self energy \eqref{B_310}. The other self-consistent equations remain
unchanged. The grand thermodynamic potential \eqref{B_130} is transformed into
\begin{widetext}
\begin{equation}
\begin{split}
\Omega = &-L^d \int \frac{d^dk}{(2\pi)^d} \ \frac{1}{\beta}
\sum_{\omega_n} \text{Tr} \bigl\{ -\ln[ G(\textbf{k},\omega_n) ] +
[ G_1(\textbf{k},\omega_n)^{-1} G(\textbf{k},\omega_n) - 1 ] \bigr\} \\
&- L^d \frac{\vert\Delta\vert^2}{g_0}
\ +\ \frac{1}{2}\, L^d \int \frac{d^dK}{(2\pi)^d} \ \frac{1}{\beta} \sum_{\Omega_n}
\text{Tr} \bigl\{ - \ln[ \Gamma(\textbf{K},\Omega_n)/g_0 ] \bigr\} \ . \\
\end{split}
\label{B_430}
\end{equation}
For a combination of the internal energy \eqref{B_200} and the particle number \eqref{B_220} we 
obtain the formula
\begin{equation}
\begin{split}
U - \mu N = &-L^d \int \frac{d^dk}{(2\pi)^d} \ \frac{1}{\beta} \sum_{\omega_n} \text{Tr} \bigl\{ 
[ G_1(\textbf{k},\omega_n)^{-1} + i\hbar\omega_n ] G(\textbf{k},\omega_n) \bigr\} \\
&- L^d \frac{\vert\Delta\vert^2}{g_0}
\ -\ \frac{1}{2}\, L^d \int \frac{d^dK}{(2\pi)^d} \ \frac{1}{\beta} \sum_{\Omega_n}
\text{Tr} \bigl\{ [ \Gamma(\textbf{K},\Omega_n)/g_0 - 1 ] \bigr\} \ . \\
\end{split}
\label{B_440}
\end{equation}
The entropy \eqref{B_174} depends neither on $G_0$ nor on $G_1$. We transform the 
formula into
\begin{equation}
\begin{split}
S = &\beta\, L^d \int \frac{d^dk}{(2\pi)^d} \ \frac{1}{\beta}
\sum_{\omega_n} \text{Tr} \bigl\{ -\ln[ G(\textbf{k},\omega_n) ] +
[ -i\hbar\omega_n G(\textbf{k},\omega_n) - 1 ] \bigr\} \\
&- \frac{1}{2}\, \beta\, L^d \int \frac{d^dK}{(2\pi)^d} \ \frac{1}{\beta}
\sum_{\Omega_n} \text{Tr} \bigl\{ - \ln[ \Gamma(\textbf{K},\Omega_n) / g_0 ]
+ [ \Gamma(\textbf{K},\Omega_n) / g_0 - 1 ] \bigr\} \ . \\
\end{split}
\label{B_450}
\end{equation}
In the above three formulas we have simplified the terms involving the vertex function $\Gamma$
by using the Bethe-Salpeter equation \eqref{B_335}. The grand thermodynamic potential \eqref{B_430}
was derived using the formalism of Luttinger and Ward \cite{LW60} while the other two quantities
\eqref{B_440} and \eqref{B_450} were derived using the formalism of DeDominicis and Martin 
\cite{DM64}. It is now not hard to see that the above expressions obey the thermodynamic relation
\begin{equation}
\Omega = U - T\,S - \mu\, N 
\label{B_455}
\end{equation}
which explicitely shows that both formalisms are indeed equivalent yielding the same results 
for all thermodynamic potentials in self-consistent ladder approximation provided $G$ and $\Gamma$ 
satisfy the appropriate stationarity equations.

\subsection{Mean-field approximation}
\label{subsection_2F}

If we insert $G_1$ for the matrix Green function $G$ and neglect all terms containing the 
vertex function $\Gamma$ we obtain the thermodynamic potentials in mean-field approximation. 
In particular the mean-field grand thermodynamic potential is given by 
\begin{equation}
\Omega_1 = -L^d \int \frac{d^dk}{(2\pi)^d} \ \frac{1}{\beta}
\sum_{\omega_n} \text{Tr} \bigl\{ -\ln[ G_1(\textbf{k},\omega_n) ] \bigr\} 
- L^d \frac{\vert\Delta\vert^2}{g_0} \ ,
\label{B_460}
\end{equation}
while the mean-field formula for the combination of the internal energy and the particle number are
\begin{equation}
U_1 - \mu N_1 = -L^d \int \frac{d^dk}{(2\pi)^d} \ \frac{1}{\beta} \sum_{\omega_n} \text{Tr} 
\bigl\{ [ G_1(\textbf{k},\omega_n)^{-1} + i\hbar\omega_n ] G_1(\textbf{k},\omega_n) \bigr\} 
- L^d \frac{\vert\Delta\vert^2}{g_0} \ ,
\label{B_470}
\end{equation}
and the mean-field entropy is
\begin{equation}
S_1 = \beta\, L^d \int \frac{d^dk}{(2\pi)^d} \ \frac{1}{\beta}
\sum_{\omega_n} \text{Tr} \bigl\{ -\ln[ G_1(\textbf{k},\omega_n) ] +
[ -i\hbar\omega_n G_1(\textbf{k},\omega_n) - 1 ] \bigr\} \ .
\label{B_480}
\end{equation}

In order to obtain finite results, we must define the sums over the Matsubara frequencies
as described in Appendix \ref{appendix_B}. The sums can be evaluated explicitly. This yields
\begin{eqnarray}
\Omega_1 &=& E_0 - \frac{1}{\beta}\, 2\, L^d \int \frac{d^dk}{(2\pi)^d} 
\ \ln[1 + \exp(-\beta (E_\textbf{k} - \mu)] \ , 
\label{B_490} \\
U_1 - \mu N_1 &=& E_0 + 2\, L^d \int \frac{d^dk}{(2\pi)^d} \  \left(E_\textbf{k} - \mu\right)\, n_\textbf{k} \ ,
\label{B_500} \\
S_1 &=& -2\, L^d \int \frac{d^dk}{(2\pi)^d} \bigl\{ (1-n_\textbf{k}) \ln(1-n_\textbf{k})
+ n_\textbf{k} \ln n_\textbf{k} \bigr\} \ ,
\label{B_510}
\end{eqnarray}
which are the well known results of a BCS variational Ansatz for arbitrary coupling where
\begin{equation}
E_0 = -2 \, L^d \int \frac{d^dk}{(2\pi)^d} \ \frac{1}{2} \bigl[ (E_\textbf{k} -\mu) 
- (\varepsilon_\textbf{k} - \mu) \bigr] \ - \  L^d \, \frac{\vert\Delta\vert^2}{g_0}
\label{B_520}
\end{equation}
\end{widetext}
is an energy constant which after renormalization $g_0 \rightarrow g$ (see \eqref{B_620}) 
reduces to the BCS condensation energy.
Here  $E_\textbf{k}$ is the spectrum of the quasiparticles, defined by
\begin{equation}
(E_\textbf{k} - \mu) = [ (\varepsilon_\textbf{k} - \mu)^2 + \vert\Delta\vert^2 ]^{1/2} \ ,
\label{B_530}
\end{equation}
and $n_\textbf{k}$ denotes the Fermi distribution function of the quasiparticles 
\begin{equation}
n_\textbf{k} = 1 / [\exp(\beta (E_\textbf{k} -\mu))+1] \ .
\label{B_540}
\end{equation}
We find that the regularization of the Matsubara-fre\-quency sums described in Appendix
\ref{appendix_B} affects only the energy constant $E_0$. The regularization has been
chosen such that for zero interaction the results for the ideal Fermi gas are obtained
which implies $E_0=0$. The other terms in \eqref{B_490}-\eqref{B_510} are not affected
by the regularization.

\subsection{Beyond mean-field}
\label{subsection_2G}
In the mean-field approximation, the formation and condensation of fermion pairs 
occur at the same temperature. This is the well known BCS scenario, which is 
perfectly captured by the exact solution of the reduced BCS-Hamiltonian. Formally, 
this solution can easily be extended to arbitrary coupling strengths \cite{Dukelsky}. 
At zero temperature, it provides a smooth crossover from the BCS groundstate
of highly overlapping pairs to a perfect Bose-Einstein condensate at infinite
coupling, similar to the variational Ansatz of Eagles and Leggett \cite{Eagles,Leggett}.
At finite temperature, however, superfluidity in this model is destroyed by
fermionic excitations, namely the breakup of pairs. The critical temperature
is therefore of the same order as the pairing gap at zero temperature,
consistent with the well known BCS relation $2\Delta_0/T_c=3.52$ in weak coupling.
Clearly, such a picture is appropriate for weak coupling, where 
the transition to superfluidity is driven by the gain in {\it potential} energy
associated with pair formation. By contrast, for sufficiently strong interactions, 
the superfluid to normal transition is instead driven by a gain in {\it kinetic} 
energy, associated with the condensation of pre-formed pairs rather than their 
thermal breakup.  The critical temperature is then of the order of the degeneracy
temperature of the gas and thus is completely unrelated to the pair binding energy. 
For a proper description of the BCS-BEC crossover at finite temperature and
arbitrary coupling,  we therefore need to go beyond mean-field, including 
excitations, which drive the superfluid order parameter to zero without destroying 
the bound pairs altogether. On a formal level, this is accomplished by  
the nontrivial wave-vector and frequency dependent term $G\Gamma$ in the 
exact fermion self energy, as given in \eqref{B_310}. 

For the numerical calculation we decompose the thermodynamic potentials into a mean-field part 
and a correction term according to $\Omega = \Omega_1 + \Delta\Omega$, $S = S_1 + \Delta S$, etc..
The mean-field contributions have been derived in the previous subsection. While in 
these contributions the Matsubara-frequency sums have been performed explicitly, the 
integrals over the wave vector remain and must be evaluated numerically. By subtracting 
the mean-field formulas \eqref{B_460}-\eqref{B_480} from the general formulas 
\eqref{B_430}-\eqref{B_450} we obtain the correction for the grand thermodynamic potential
\begin{widetext}
\begin{equation}
\begin{split}
\Delta\Omega = &-L^d \int \frac{d^dk}{(2\pi)^d} \ \frac{1}{\beta}
\sum_{\omega_n} \text{Tr} \bigl\{ -\ln[ G_1(\textbf{k},\omega_n)^{-1} G(\textbf{k},\omega_n) ] +
[ G_1(\textbf{k},\omega_n)^{-1} G(\textbf{k},\omega_n) - 1 ] \bigr\} \\
&+\frac{1}{2}\, L^d \int \frac{d^dK}{(2\pi)^d} \ \frac{1}{\beta} \sum_{\Omega_n}
\text{Tr} \bigl\{ - \ln[ \Gamma(\textbf{K},\Omega_n)/g_0 ] \bigr\} \ , \\
\end{split}
\label{B_590}
\end{equation}
the correction for the combination of the internal energy and the particle number
\begin{equation}
\begin{split}
\Delta U - \mu \Delta N = &-L^d \int \frac{d^dk}{(2\pi)^d} 
\ \frac{1}{\beta} \sum_{\omega_n} \text{Tr} \bigl\{ 
[ G_1(\textbf{k},\omega_n)^{-1} + i\hbar\omega_n ] 
[ G(\textbf{k},\omega_n) -  G_1(\textbf{k},\omega_n) ] \bigr\} \\
&- \frac{1}{2}\, L^d \int \frac{d^dK}{(2\pi)^d} \ \frac{1}{\beta} \sum_{\Omega_n}
\text{Tr} \bigl\{ [ \Gamma(\textbf{K},\Omega_n)/g_0 - 1 ] \bigr\} \ , \\
\end{split}
\label{B_600}
\end{equation}
and the correction for the entropy
\begin{equation}
\begin{split}
\Delta S =& \beta\, L^d \int \frac{d^dk}{(2\pi)^d} \ \frac{1}{\beta}
\sum_{\omega_n} \text{Tr} \bigl\{ -\ln[ G_1(\textbf{k},\omega_n)^{-1} G(\textbf{k},\omega_n) ] +
( -i\hbar\omega_n ) [ G(\textbf{k},\omega_n) - G_1(\textbf{k},\omega_n) ] \bigr\} \\
&- \frac{1}{2}\, \beta\, L^d \int \frac{d^dK}{(2\pi)^d} \ \frac{1}{\beta}
\sum_{\Omega_n} \text{Tr} \bigl\{ - \ln[ \Gamma(\textbf{K},\Omega_n) / g_0 ]
+ [ \Gamma(\textbf{K},\Omega_n) / g_0 - 1 ] \bigr\} \ . \\
\end{split}
\label{B_610}
\end{equation}
In formulas \eqref{B_590}-\eqref{B_610} the sums over the fermionic Matsubara 
frequencies $\omega_n$ converge so that the regularization of Appendix \ref{appendix_B} is not
needed. However, the sums over the bosonic Matsubara frequencies $\Omega_n$
are not well defined and must be regularized. Thus, for a numerical evaluation the formulas 
\eqref{B_590}-\eqref{B_610} must be transformed further, which will be done in the next
subsection.

\subsection{Renormalization of the thermodynamic potentials}
\label{subsection_2H}

Since the interaction has zero range, the interaction strength $g_0$ must be renormalized
and replaced by the scattering amplitude $g$ according to \eqref{B_030}. In a first
step we renormalize the mean-field formulas of the thermodynamic potentials. 
In \eqref{B_490}-\eqref{B_510} the interaction strength $g_0$ does not occur explicitly.
The integrals are thus finite and no renormalization is needed for these formulas.
However, the condensation energy  \eqref{B_520} contains two infinite
terms, a divergent integral and the last term with the infinite factor $1/g_0$, which 
compensate each other. By renormalizing the interaction strength we obtain
\begin{equation}
E_0 = -2 \, L^d \int \frac{d^dk}{(2\pi)^d} \ \frac{1}{2} \Bigl[ (E_\textbf{k} -\mu)
- (\varepsilon_\textbf{k} - \mu) - \frac{\vert\Delta\vert^2}{2 \varepsilon_\textbf{k} } \Bigr] 
\ - \ L^d \, \frac{\vert\Delta\vert^2}{g} \\
\label{B_620}
\end{equation}
where both the integral and the last term are now separately finite. Note that  
the wave vector integrals in \eqref{B_490}-\eqref{B_510} and in \eqref{B_620} are finite 
in any spatial dimension $d$ with $2<d<4$.

In a second step we renormalize the correction formulas. In the correction of the grand
thermodynamic potential \eqref{B_590} we decompose $\ln[ \Gamma(\textbf{K},\Omega_n)/g_0 ]
= \ln[ \Gamma(\textbf{K},\Omega_n)/g ] + \ln[g/g_0]$. The separated term
$\ln[g/g_0]$ can be neglected because it does not depend on the Matsubara frequencies
$\Omega_n$. Following the arguments of Appendix \ref{appendix_B} the Matsubara-frequency sum
of this term is zero. Thus, for the correction of the grand thermodynamic potential we obtain
the renormalized formula
\begin{equation}
\begin{split}
\Delta\Omega = &-L^d \int \frac{d^dk}{(2\pi)^d} \ \frac{1}{\beta}
\sum_{\omega_n} \text{Tr} \bigl\{ -\ln[ G_1(\textbf{k},\omega_n)^{-1} G(\textbf{k},\omega_n) ] +
[ G_1(\textbf{k},\omega_n)^{-1} G(\textbf{k},\omega_n) - 1 ] \bigr\} \\
&+\frac{1}{2}\, L^d \int \frac{d^dK}{(2\pi)^d} \ \frac{1}{\beta} \sum_{\Omega_n}
\text{Tr} \bigl\{ - \ln[ \Gamma(\textbf{K},\Omega_n)/g ] \bigr\} \ . \\
\end{split}
\label{B_630}
\end{equation}
Both terms of this formula are now finite. However, Eq.\ \eqref{X_140} is needed to evaluate 
the second term.

In correction \eqref{B_600} the second term must be renormalized. This can be achieved by the 
following sequence of equations
\begin{equation}
\begin{split}
&- L^d \int \frac{d^dK}{(2\pi)^d} \ \frac{1}{\beta} \sum_{\Omega_n}
\text{Tr} \bigl\{ [ \Gamma(\textbf{K},\Omega_n)/g_0 - 1 ] \bigr\} 
= L^d \int \frac{d^dK}{(2\pi)^d} \ \frac{1}{\beta} \sum_{\Omega_n}
\text{Tr} \bigl\{ \Gamma(\textbf{K},\Omega_n) \chi(\textbf{K},\Omega_n) \bigr\} \\ 
&= L^d \int \frac{d^dk}{(2\pi)^d} \ \frac{1}{\beta} \sum_{\omega_n} 
\text{Tr} \bigl\{ \tilde \Sigma(\textbf{k},\omega_n) G(\textbf{k},\omega_n) ] \bigr\}
= L^d \int \frac{d^dk}{(2\pi)^d} \ \frac{1}{\beta} \sum_{\omega_n} 
\text{Tr} \bigl\{ G_1(\textbf{k},\omega_n)^{-1} [ G(\textbf{k},\omega_n) 
- G_1(\textbf{k},\omega_n) ] \bigr\} \ . \\
\end{split}
\label{B_640}
\end{equation}
First, by using the Bethe-Salpeter equation \eqref{B_335} we write the integrand as a
product of the matrix vertex function $\Gamma$ and the matrix pair propagator $\chi$. 
Secondly, we express $\chi$ in terms of the matrix Green function $G$ by \eqref{B_150},
interchange the orders of the integrals and sums, and combine $\Gamma$ with one of the
$G$ into the self energy $\tilde \Sigma$ by \eqref{B_420}. Finally, we replace $\tilde \Sigma$
in favor of $G$ and $G_1$ by using the Dyson equation \eqref{B_410}. The bosonic integral
and sum are transformed into a fermionic integral and sum. Hence, the second term of 
\eqref{B_600} can be combined with the first term. As a result we finally obtain
\begin{equation}
\Delta U - \mu \Delta N = -\frac{1}{2}\, L^d \int \frac{d^dk}{(2\pi)^d} 
\ \frac{1}{\beta} \sum_{\omega_n} \text{Tr} \bigl\{ 
[ G_1(\textbf{k},\omega_n)^{-1} + 2i\hbar\omega_n ] 
[ G(\textbf{k},\omega_n) -  G_1(\textbf{k},\omega_n) ] \bigr\} \ .
\label{B_650}
\end{equation}
By considering \eqref{B_400} we explicitly prove
\begin{equation}
G_1(\textbf{k},\omega_n)^{-1} + 2i\hbar\omega_n = G_1(\textbf{k},-\omega_n)^{-1} \ .
\label{B_660}
\end{equation}
Consequently, for the correction of the combination of the internal energy and the 
particle number we obtain the compact formula
\begin{equation}
\Delta U - \mu \Delta N = -\frac{1}{2}\, L^d \int \frac{d^dk}{(2\pi)^d} 
\ \frac{1}{\beta} \sum_{\omega_n} \text{Tr} \bigl\{ G_1(\textbf{k},-\omega_n)^{-1} 
[ G(\textbf{k},\omega_n) -  G_1(\textbf{k},\omega_n) ] \bigr\} 
\label{B_670}
\end{equation}
which is essential for a stable numerical evaluation of the correction term.
The Matsubara-frequency sum is evaluated by using \eqref{X_130}. The wave-vector integral
is finite.

The correction of the entropy \eqref{B_610} is renormalized in an analogous way. Alternatively,
we use the thermodynamic relation \eqref{B_455}. As a result we obtain
\begin{equation}
\begin{split}
\Delta S=& \beta\, L^d \int \frac{d^dk}{(2\pi)^d} \ \frac{1}{\beta}
\sum_{\omega_n} \text{Tr} \bigl\{ -\ln[ G_1(\textbf{k},\omega_n)^{-1} G(\textbf{k},\omega_n) ] +
[ G_1(\textbf{k},\omega_n)^{-1} G(\textbf{k},\omega_n) - 1 ] \bigr\} \\
&-\frac{1}{2}\, \beta\, L^d \int \frac{d^dk}{(2\pi)^d}
\ \frac{1}{\beta} \sum_{\omega_n} \text{Tr} \bigl\{ G_1(\textbf{k},-\omega_n)^{-1}
[ G(\textbf{k},\omega_n) -  G_1(\textbf{k},\omega_n) ] \bigr\} \\
&- \frac{1}{2}\, \beta\, L^d \int \frac{d^dK}{(2\pi)^d} \ \frac{1}{\beta}
\sum_{\Omega_n} \text{Tr} \bigl\{ - \ln[ \Gamma(\textbf{K},\Omega_n) / g] \bigr\} \ . \\
\end{split}
\label{B_680}
\end{equation}
\end{widetext}

The final results are the mean-field formulas \eqref{B_490}-\eqref{B_510} together with
\eqref{B_620} and the correction formulas \eqref{B_630}, \eqref{B_670}, and \eqref{B_680}.
In these formulas each term by itself is finite. Eventually, the thermodynamic potentials
are obtained by adding the terms together according to $\Omega = \Omega_1 + \Delta\Omega$, 
$S = S_1 + \Delta S$, etc..

\subsection{Symmetry breaking and Thouless criterion}
\label{subsection_2I}

The interacting fermion system is invariant under the symmetry transformation
\begin{equation}
\psi^{\ }_\sigma(\textbf{r}) \rightarrow 
e^{i\lambda} \psi^{\ }_\sigma(\textbf{r}) \ , \quad
\psi^+_\sigma(\textbf{r}) \rightarrow 
e^{-i\lambda} \psi^+_\sigma(\textbf{r})
\label{B_690}
\end{equation}
which is related to a global change of  phase of the fermion fields by $\lambda$.
The superfluid phase breaks this symmetry since the order parameter $\Delta$ 
is transformed as $\Delta \rightarrow e^{2i\lambda} \Delta$. Clearly, 
however, the thermodynamic potentials must remain invariant under 
a global change of the phase both in the 
normal and in the superfluid state. In the superfluid, the free energy increase
associated with a {\it slowly varying} phase $\lambda(\textbf{r})$ vanishes like 
$(\nabla\lambda)^2$. By Goldstone' s theorem, this implies the existence of
modes whose energy vanishes in the long wave-length limit. For a neutral superfluid, 
this is the well known Bogoliubov-Anderson mode. It has a sound like 
dispersion $\omega(k)=c k$ and is physically related to fluctuations 
of the phase of the order parameter.

In technical terms, the existence of zero-energy collective modes can be derived 
from Ward identities related to the symmetry transformation. By considering the 
grand thermodynamic potential $\Omega[G]$, in Ref.\ \onlinecite{Haussmann3} 
the Ward identity
\begin{equation}
\sum_{Y Y^\prime} \Gamma^{-1}_{X X^\prime,Y^\prime Y} 
\, \delta_\lambda \Sigma^{\ }_{Y Y^\prime} = 0
\label{B_700}
\end{equation}
has been derived (see (2.57) in Ref.\ \onlinecite{Haussmann3}). Here $\delta_\lambda
\Sigma_{X X^\prime}$ is the variation of the self energy under the transformation
\eqref{B_690} with an infinitesimal phase change $\delta\lambda$. This quantity may be
interpreted as the generalized order parameter of the system. On the other hand,
$\Gamma^{-1}_{X X^\prime,Y Y^\prime}$ is the inverse vertex function. For a short-hand
notation the indices $X$, $X^\prime$ and $Y$, $Y^\prime$ are used, which represent a 
combination $X=(\textbf{r},\tau,\sigma,\alpha)$ of the space variable $\textbf{r}$, 
the imaginary time $\tau$, the spin index $\sigma$, and the Nambu index $\alpha$. 
According to \eqref{B_700} the inverse vertex function 
$\Gamma^{-1}$ may be interpreted as a linear operator which acts on the order parameter 
$\delta_\lambda \Sigma$. In the superfluid state the order parameter is 
nonzero so that the inverse vertex function must have a zero eigenvalue, which is related 
to a zero energy collective mode. For superfluid fermion systems this fact is known as 
the Thouless criterion \cite{Th60}.

The Ward identity \eqref{B_700} has been derived for the exact theory. However, our present 
crossover theory is an approximation, based on a certain truncation of the exact
functional which enters either into the Luttinger-Ward or the DeDominicis-Martin formalism.
In general, such a truncated functional will not obey the Ward identity. Indeed, we find 
that our inverse vertex function $\Gamma^{-1}_{\alpha\alpha^\prime}(\textbf{K},\Omega_n)$ 
obeys instead the equation
\begin{equation}
\sum_{\alpha^\prime} \Gamma^{-1}_{\alpha\alpha^\prime}(\textbf{K}=\textbf{0},\Omega_n=0)
\, \Delta_{\alpha^\prime} = {\cal O}( \vert\Delta\vert^3 )
\label{B_710}
\end{equation}
where $(\Delta_\alpha)=(\Delta, \Delta^*)$ (see (3.56) in 
Ref.\ \onlinecite{Haussmann3}). Taking the longitudinal part, this equation
correctly describes the smooth evolution from a Ginzburg-Landau type
description of  weak coupling BCS superfluids to a Gross-Pitaevskii like theory 
of a dilute, repulsive Bose gas \cite{Haussmann3}. The transverse part,
however, also gives a finite value on the right hand side of \eqref{B_710}
in the limit $\textbf{K}=\textbf{0}$ and $\Omega_n=0$, 
thus violating the Ward identity by terms of order $\vert\Delta\vert^3$.
As a result, the Thouless criterion is violated and there is no proper Bogoliubov-Anderson
mode in our approach without a further modification (see below).

Unfortunately, the violation of the Goldstone theorem for continuous symmetries
is a general property of conserving approximations based on the Luttinger-Ward formalism.
This problem has been known for a long time for superfluid Bose systems
\cite{HM65} and is sometimes referred to as the 'conserving-gapless dichotomy' \cite{Strinati04,Kita06} 
in the literature. For the exact theory, a Ward identity holds for the inverse matrix boson Green 
function $G_{B}$, which reads
\begin{equation}
\sum_{\alpha^\prime} G^{-1}_{B,\alpha\alpha^\prime}(\textbf{K}=\textbf{0},\Omega_n=0) 
\, \Psi_{B,\alpha^\prime} = 0 \ .
\label{B_720}
\end{equation}
In the superfluid state the inverse matrix boson 
Green function has a vanishing eigenvalue. For superfluid boson systems this 
is known as the Hugenholtz-Pines theorem \cite{HP59}. Conserving 
approximations, however, violate the Hugenholtz-Pines theorem. For example, this is true 
already for the lowest approximation, the well known Hartree-Fock-Bogoliubov theory.

In our fermion system for strong attractive interactions $v=1/k_F a \gg 1$, the fermions
are bound into pairs.  These pairs form a Bose system with an effective repulsive interaction
which, for a dilute system, is described by the exact scattering length
$a_{dd}\approx 0.60\, a$ of the four particle problem associated with dimer-dimer scattering.
In the strong coupling limit, therefore,  our crossover theory for interacting fermions 
must converge to an effective theory of repulsively interacting bosons, where both theories are based 
on the Luttinger-Ward formalism. From the analytical arguments  in 
Refs.\ \onlinecite{Haussmann1,Haussmann3} and also from our numerical calculations, 
we find that the crossover theory
converges to the Hartree-Fock-Bogoliubov theory quickly for interactions $v=1/k_F a > 2$.
The boson order parameter $\Psi_B$ and the matrix boson 
Green function $G_B(\textbf{K},\Omega_n)$ can be identified with the order parameter 
$\Delta$ and the vertex function $\Gamma(\textbf{K},\Omega_n)$ according to \cite{Haussmann1,
Haussmann3}
\begin{eqnarray}
\Psi_B &=& \pm i [8\pi\varepsilon_b^2 a^3]^{-1/2} \Delta \ ,
\label{B_730} \\
G_B(\textbf{K},\Omega_n) &=& - [8\pi\varepsilon_b^2 a^3]^{-1}\Gamma(\textbf{K},\Omega_n) \ . 
\label{B_740}
\end{eqnarray}
The validity of these relations requires both strong coupling, but also low frequencies 
and momenta. Indeed, it is only at low energies where the composite particles behave like bosons.
At higher frequencies or momenta, the composite nature of the pairs becomes visible. 
This becomes evident, for instance, in the different behaviour $G_B\sim\Omega_n^{-1}$ of a 
Bose Green function at large frequencies compared to that of the vertex function, 
which behaves like $\Gamma\sim\Omega_n^{-1/2}$ as a result of the two particle 
continuum associated with broken fermion pairs. Clearly, at large coupling constants
$v\gg 1$, this continuum moves up to very large frequencies of the order of the 
binding energy $\epsilon_b\sim v^2$.

We conclude that the violation of the Thouless criterion in our crossover theory is 
related to the violation of the Hugenholtz-Pines theorem in the Hartree-Fock-Bogoliubov
theory for bosons to which our Luttinger-Ward formulation of the fermionic many-body 
problem converges at large coupling. In the following section, it will be shown that 
this problem may be solved by an appropriate modification of the coupling constant.
In this manner, a self-consistent formulation of the many-body problem is possible
which obeys Goldstone's theorem and thus provides a correct description of both 
fermionic and collective, bosonic excitations along the BCS-BEC crossover.

\subsection{Modified coupling and gapless Bogoliubov-Anderson mode}
\label{subsection_2J}

In the following, our aim is to modify the theory in a way which is consistent with the 
Thouless criterion, giving rise to a gapless Goldstone mode in the whole regime of 
coupling strengths. If we require the Thouless criterion 
\begin{equation}
\sum_{\alpha^\prime} \Gamma^{-1}_{\alpha\alpha^\prime}(\textbf{K}=\textbf{0},\Omega_n=0)
\, \Delta_{\alpha^\prime} = 0
\label{B_750}
\end{equation}
a further equation will be added to the self-consistent equations for the Green and vertex
functions in Subsec.\ \ref{subsection_2D}. However, then another equation must be discarded
or a further parameter must be introduced. We find that the Bethe-Salpeter equation 
\eqref{B_350} and the order-parameter equation \eqref{B_390} can not be satisfied
together, if \eqref{B_750} is required. For this reason we modify the theory by
introducing a modified scattering amplitude $g_\text{mod}$, which is
determined by the modified order-parameter equation
\begin{equation}
\Delta = g_\text{mod} \int \frac{d^dk}{(2\pi)^d} 
\Bigl[ {\cal F}(\textbf{k},\tau=0) + \Delta\, \frac{m}{\hbar^2 \textbf{k}^2} \Bigr] \ .
\label{B_760}
\end{equation}
We have solved the self-consistent equations together with the Thouless criterion
\eqref{B_750} and the modified order-parameter equation \eqref{B_760}. The numerical 
effort is much less for the modified theory than for the original theory.
Since the scattering amplitude $g$ is related to the scattering length $a$,
we obtain a modified dimensionless interaction strength $v_\text{mod}=1/k_F a_\text{mod}$.
We find a difference $\delta v_\text{mod} = v_\text{mod} - v$ in the range between
$0.0$ and $-0.1$.

In order to obtain a consistent theory, we must check that the modification is 
compatible with the Luttinger-Ward formalism. We must find a modified grand thermodynamic 
potential $\Omega_\text{mod}[G]$, so that the condition for stationarity \eqref{B_120} yields 
the self-consistent equations with the modified order-parameter equation \eqref{B_760}.
For this purpose we consider the second term of \eqref{B_130} which reads
\begin{equation}
\Omega_0[G] = L^d\, g_0 \vert {\cal F}(\textbf{0},0) \vert^2 
= L^d\, \vert\Delta\vert^2 / g_0 \ .
\label{B_770}
\end{equation}
We replace this term by the modified term 
\begin{equation}
\Omega_{0,\text{mod}}[G] = L^d\, \vert\Delta\vert^2 / \tilde g_{0,\text{mod}}(\vert\Delta\vert) 
\label{B_780}
\end{equation}
where
\begin{equation}
\Delta = g_{0,\text{mod}}(\vert\Delta\vert) \, {\cal F}(\textbf{0},0)
\label{B_790}
\end{equation}
is the modified order-parameter equation. The modified interaction strengths 
$\tilde g_{0,\text{mod}}= \tilde g_{0,\text{mod}}(\vert\Delta\vert)$ and 
$g_{0,\text{mod}}= g_{0,\text{mod}}(\vert\Delta\vert)$ depend on the order parameter
$\vert\Delta\vert$, are not equal, and differ from the bare interaction strength $g_0$. 
In order to apply the stationarity condition \eqref{B_120} we must consider the variation of
\eqref{B_780} with respect to $G$. Since the modified parameter 
$\tilde g_{0,\text{mod}}(\vert\Delta\vert)$ depends implicitly on $G$ via \eqref{B_790}, the
chain rule of differential calculus must be applied. Eventually, the variation of
\eqref{B_780} must have the form
\begin{equation}
\begin{split}
\delta \Omega_{0,\text{mod}}[G] &= L^d \int \frac{d^dk}{(2\pi)^d} \, \frac{1}{\beta} \sum_{\omega_n}
\text{Tr} \bigl\{ \Sigma_1 \, \delta G(\textbf{k},\omega_n) \bigr\} \\
&= L^d [ \Delta\, \delta {\cal F}(\textbf{0},0)^* + \Delta^* \, \delta {\cal F}(\textbf{0},0) ] \ .
\end{split}
\label{B_795}
\end{equation}
By comparing the resulting terms with \eqref{B_795}, we obtain the differential equation
\begin{equation}
\frac{\partial}{\partial \vert\Delta\vert}
\frac{\vert\Delta\vert^2}{\tilde g_{0,\text{mod}}(\vert\Delta\vert) } =
2 \vert\Delta\vert \frac{\partial}{\partial \vert\Delta\vert}
\frac{\vert\Delta\vert}{g_{0,\text{mod}}(\vert\Delta\vert) } \ . 
\label{B_800}
\end{equation}
On the other hand, Eq.\ \eqref{B_710} implies that the Thouless criterion holds without 
modification if $\vert\Delta\vert=0$. Thus, we find
\begin{equation}
\tilde g_{0,\text{mod}} = g_{0,\text{mod}} = g_0 \quad \text{for $\vert\Delta\vert=0$} 
\label{B_810}
\end{equation}
which is an initial condition for \eqref{B_800}. Eq.\ \eqref{B_800} can be integrated
together with \eqref{B_810}. We obtain
\begin{equation}
\frac{1}{\tilde g_{0,\text{mod}}(\vert\Delta\vert)} = 
\frac{2}{g_{0,\text{mod}}(\vert\Delta\vert)}
- \frac{1}{\vert\Delta\vert^2} \int_0^{\vert\Delta\vert}
\frac{2\vert\Delta^\prime\vert\, d\vert\Delta^\prime\vert}
{g_{0,\text{mod}}(\vert\Delta^\prime\vert)} \ .
\label{B_820}
\end{equation}
The thermodynamic state of the interacting fermion system
in the superfluid state is therefore determined by three 
parameters. We may choose the order parameter $\vert\Delta\vert$, the chemical 
potential $\mu$, and the interaction strength $g_0$ for these parameters. Hence, the
modified interaction strengths $g_{0,\text{mod}}= g_{0,\text{mod}}(\vert\Delta\vert,\mu,g_0)$
and $\tilde g_{0,\text{mod}}= \tilde g_{0,\text{mod}}(\vert\Delta\vert,\mu,g_0)$ are functions
of these parameters. While $g_{0,\text{mod}}(\vert\Delta\vert,\mu,g_0)$ is uniquely determined
by \eqref{B_790} and the other self-consistent equations, 
$\tilde g_{0,\text{mod}}(\vert\Delta\vert,\mu,g_0)$ depends on the path in the parameter space
when the integral \eqref{B_820} is calculated. Since $g_0$ and $\mu$ are external parameters
of the theory, for a correct formulation of the modification these parameters must be kept
constant.

The modification is compatible also with the DeDo\-minicis-Martin formalism. In
this case the internal energy $U[G,\Gamma]$ includes the term \eqref{B_770}
which must be modified according to \eqref{B_780}. The modification of the
coupling constant $g_0$ described by \eqref{B_800}-\eqref{B_820} is derived
in an analogous way.

\begin{figure}[t]
\includegraphics[width=0.4\textwidth]{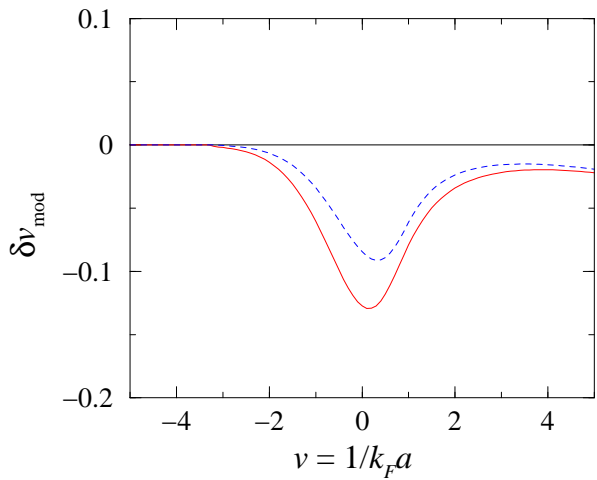}
\caption{(Color online) Modifications of the dimensionless interaction parameter: red solid curve shows 
$\delta v_\text{mod}$ and blue dashed curve depicts $\delta \tilde v_\text{mod}$ as a 
function of $v$ for $T=0$ and $\vert\Delta\vert = \vert\Delta_0\vert$.}
\label{Fig_delta_v}
\end{figure}

Eqs.\ \eqref{B_770}-\eqref{B_820} describe the modification of the crossover theory
in terms of the bare interaction parameters $g_0$, $g_{0,\text{mod}}$, and 
$\tilde g_{0,\text{mod}}$. A renormalized version of the modification is obtained, if we
replace the bare parameters by the renormalized scattering amplitudes $g$,
$g_\text{mod}$, and $\tilde g_\text{mod}$ according to \eqref{B_030}. 
Eqs.\ \eqref{B_800}-\eqref{B_820} are valid also for the renormalized scattering 
amplitudes without changes. From \eqref{B_820} we obtain
\begin{equation}
\frac{1}{\tilde g_\text{mod}(\vert\Delta\vert)} = 
\frac{2}{g_\text{mod}(\vert\Delta\vert)}
- \frac{1}{\vert\Delta\vert^2} \int_0^{\vert\Delta\vert}
\frac{2\vert\Delta^\prime\vert\, d\vert\Delta^\prime\vert}
{g_\text{mod}(\vert\Delta^\prime\vert)} \ .
\label{B_830}
\end{equation}
The renormalized modified order-parameter equation is defined by \eqref{B_760}.
In order to obtain the modified formulas of the renormalized thermodynamic potentials
in Subsec.\ \ref{subsection_2H} only a single change is needed. We must replace the
energy constant \eqref{B_620} by
\begin{equation}
\begin{split}
E_{0,\text{mod}} = &-2 \, L^d \int \frac{d^dk}{(2\pi)^d} \ \frac{1}{2} \Bigl[ 
(E_\textbf{k} - \mu)  - (\varepsilon_\textbf{k} - \mu) 
- \frac{\vert\Delta\vert}{2 \varepsilon_\textbf{k} } \Bigr] \\
&+ \ L^d \, \vert\Delta\vert^2 \, ( \tilde g_\text{mod}^{-1} - 2\, g^{-1} ) \ . \\
\end{split}
\label{B_840}
\end{equation}
The other formulas \eqref{B_630}, \eqref{B_670}, and \eqref{B_680} remain unchanged.
Since the renormalized scattering amplitude $g$ is related to the dimensionless
interaction parameter $v=1/k_F a$, we can transform \eqref{B_830} into a dimensionless
form. For $\delta v_\text{mod} = v_\text{mod} - v$ and $\delta \tilde v_\text{mod} =
\tilde v_\text{mod} - v$ we obtain
\begin{equation}
\begin{split}
\delta \tilde v_\text{mod}(\vert\Delta\vert/&\varepsilon_F,v) =
2\, \delta v_\text{mod}(\vert\Delta\vert/\varepsilon_F,v) \\
&- (\vert\Delta\vert/\varepsilon_F)^{-2} \int_0^{\vert\Delta\vert/\varepsilon_F}
\delta v_\text{mod}(X,v) \, 2X \, dX \ .
\end{split}
\label{B_850}
\end{equation}
While $\delta v_\text{mod}$ is obtained directly from \eqref{B_760} by solving the 
self-consistent equations, $\delta \tilde v_\text{mod}$ is obtained by evaluating the
integral in \eqref{B_850} numerically. As a result we obtain modifications which are
restricted to the interval 
\begin{equation}
-0.1 \lesssim \delta v_\text{mod} < \delta \tilde v_\text{mod} < 0 \quad
\text{for $\vert\Delta\vert > 0$} \ .
\label{B_860}
\end{equation}
In Fig.\ \ref{Fig_delta_v} the modifications $\delta v_\text{mod}$ and 
$\delta \tilde v_\text{mod}$ are shown as red solid curve and blue dashed curve, respectively, 
for $T=0$ and $\vert\Delta\vert = \vert\Delta_0\vert$. Clearly, the modifications are largest 
in the crossover region close to the unitarity point. At finite temperature for increasing $T$ 
the order parameter $\vert\Delta\vert$ and the modifications $\delta v_\text{mod}$ and 
$\delta \tilde v_\text{mod}$ decrease together. Eventually, for $\vert\Delta\vert = 0$ 
the modifications are $\delta v_\text{mod}=\delta \tilde v_\text{mod}=0$.

In the previous subsection we have argued that for strong attractive interactions the
fermions are bound into pairs. Our crossover theory for the  
interacting fermion system then converges to a Luttinger-Ward type
theory for interacting bosons. Since the modified version of
our theory obeys the Ward identity, 
its strong coupling limit necessarily leads
to a description of dilute, repulsive bosons which has the 
correct linear spectrum of excitations at low energies. It turns out,
that the limiting theory here is the Luttinger-Ward version of 
the so-called Shohno theory \cite{Sh64,ReattoStraley69} which is 
equivalent to the more well known Popov approximation. 
While we  have not been able to derive
the Shohno-Popov theory analytically from the Luttinger-Ward
functional of the original fermionic model, we find a quick convergence
numerically in all thermodynamic quantities for dimensionless  
couplings $v>2$. Considering the entropy, in particular, the 
Shohno-Popov theory gives rise to the standard expression
\begin{equation}
S = L^d \int \frac{d^dK}{(2\pi)^d} \bigl\{ (1+n^B_\textbf{K}) \ln(1+n^B_\textbf{K})
- n^B_\textbf{K} \ln n^B_\textbf{K} \bigr\}
\label{B_870}
\end{equation}
for the entropy of a noninteracting gas of bosonic quasiparticles
with the standard distribution function
\begin{equation}
n^B_\textbf{K} = 1/[ \exp(\beta[ E^B_\textbf{K} -\mu_B]) - 1] \ .
\label{B_880}
\end{equation}
The corresponding spectrum of excitation energies
\begin{equation}
E^B_\textbf{K} - \mu_B = \big[ (\hbar^2 \textbf{K}^2 /2m_B)^2 + (\hbar^2 \textbf{K}^2 /2m_B)
\, 2g_B \vert\Psi_B\vert^2 \big]^{1/2}
\label{B_890}
\end{equation}
has the well known form of a Bogoliubov spectrum with a 
temperature dependent condensate density $n_{B,0}=\vert\Psi_B\vert^2$
and a positive Bose-Bose scattering amplitude $g_B$. Within our
approximation, we have $g_B=2g$, i.e.\ the exact dimer-dimer scattering 
length $a_{dd}\approx 0.60\, a$ is replaced by its Born approximation
result $a_{dd}^{(B)}=2a$ \cite{Randeria1}. The effective mass and chemical
potential take their obvious values $m_B=2m$ and $\mu_B=2\mu+\varepsilon_b$
where $\varepsilon_b=\hbar^2/ma^2$ is the two-particle binding energy
on the BEC-side of the crossover, where $a>0$. The order parameter
$\Psi_B$ is given by \eqref{B_730}. Other thermodynamic
quantities are obtained using more complicated expressions, which 
are not presented here in detail \cite{Haussmann4}. As will be shown
explicitely in the following sections, the numerical results for the 
critical temperature or the entropy converge 
quickly to that of the Shohno-Popov theory for coupling strengths $v>2$.

\section{Numerical results}
\label{sectionNR}

Following the detailed discussion of the formalism used to describe the
thermodynamics of attractively interacting fermions at arbitrary coupling and temperature,
we now present numerical results which cover both the normal and superfluid
regime. These results require a solution of the self-consistent equations determining 
the Green function $G$ and the vertex function $\Gamma$, which are scalars above, and  
two-by-two matrices below the critical temperature. An iteration procedure is  
performed where a numerical Fourier transformation is needed to transform the functions 
back and forth. Since the Green function $G$, the vertex function $\Gamma$, and the  
related functions $\tilde\Sigma$ and $M$ are singular at small values of $\mathbf{r}$ and  
$\tau$ and also exhibit significant variation over several orders of magnitude,
the numerical Fourier transformation is quite challenging. In practice, the variables
need to  be discretized on logarithmic scales. Standard procedures like  
the fast Fourier transformation are therefore not applicable. The basic  
principles of our special numerical Fourier transformation are described in
Appendix \ref{appendix_D}.

\subsection{Critical temperature}
\label{subsection_ct}

The crucial quantity which determines the overall structure of the phase diagram is
of course the critical temperature $T_c$ for the transition to a superfluid. This
temperature is known analytically only in the extreme BCS- and BEC-limit.
In the BCS-limit $k_F|a|\to 0$, where the average
distance between the fermions is much larger than the magnitude of the
scattering length, the standard solution of the gap-equation for an attractive
pseudo-potential gives a critical temperature
\begin{equation}
T_c^{(BCS)}=\frac{8e^{\gamma_E}}{\pi e^2}\varepsilon_F\exp\bigl(-\pi/2k_F|a|\bigr)
\label{T_c/BCS}
\end{equation}
with $\gamma_E = 0.5772\ldots$ Euler's constant. $T_c^{(BCS)}$ is 
exponentially small on the characteristic scale of the Fermi energy.
Since typical Fermi temperatures in cold gases are of the order of micro-Kelvin,
the BCS-regime is in practice hardly attainable in these systems.

The leading order corrections to the BCS-result in an expansion in the  
small parameter $k_F|a|\ll 1$ have been
determined a long time ago by Gorkov and Melik-Barkhudarov
\cite{GMB61}. They arise from induced interactions, where one fermion
sees the polarization in the Fermi gas due to a second fermion. The  
density induced interaction changes the dimensionless
coupling constant $N(0)g=2k_Fa/\pi$ of the BCS-theory to  
\cite{Heiselberg00} to
\begin{equation}
g\to g+g^2N(0)\frac{1+2\ln{2}}{3}
\label{inducedinteractions}
\end{equation}
where $N(0)=mk_F/2\pi^2\hbar^2$ is the standard density of states
per spin at the Fermi energy.
Since the additional contribution to the two-body scattering amplitude
$g<0$ is positive, the induced interactions weaken the attractive 
interaction between two fermions in vacuum and lead to a 
reduction of the transition temperature by a factor
$(4e)^{-1/3}\approx 0.45$. The nonanalytic dependence of 
the BCS-transition temperature on the dimensionless coupling constant
$k_Fa$ thus give rise to a finite change in the prefactor in  
\eqref{T_c/BCS} from the BCS value $0.61$ to $0.28$, even 
though the contribution of induced interactions is of order $k_Fa$ 
compared to the bare  interaction.

\begin{figure}[t]
\includegraphics[width=0.4\textwidth]{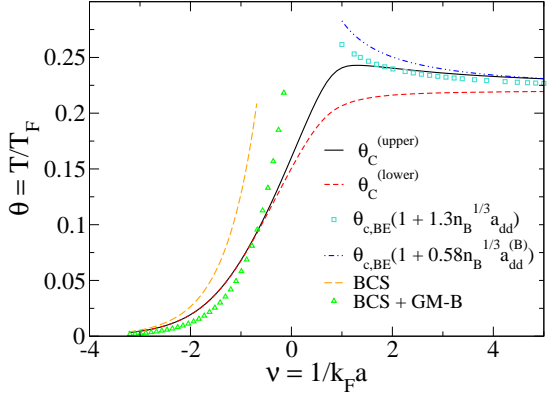}
\caption{(Color online) $ \theta_c^{(lower)}$ (red dashed line) and $\theta_c^{(upper)}$ (solid black line, 
identified as $T_c$) compared with the Shohno result (blue dotted-dashed line) with 
$a_{dd}^{(B)} = 2\, a$ and the exact (QMC) result (light-blue squares) with 
$\Delta T_c/T_{BEC} = c\,n_B^{1/3}a_{dd}$ and $c=1.31$ and $a_{dd} = 0.60\, a$.
Yellow dashed line and green triangles show the BCS result without and with 
Gorkov and Melik-Barkhudarov corrections.}
\label{Fig_phd}
\end{figure}

On the BEC-side, the zeroth order result for 
the critical temperature is obtained from the value
\begin{equation}
T_c^{(BEC)}=3.31\frac{\hbar^2n_B^{2/3}}{m_B}=0.218\,\varepsilon_F
\label{T_c/BEC}
\end{equation}
obtained for an ideal Bose gas with density $n_B=n/2$ and mass $m_B=2m$.
The leading corrections to this result arise from the residual  
interactions between the strongly bound bosonic dimers. As shown 
by Petrov \textit{et al.}\ \cite{Petrov1,Petrov2},
these interactions can be described by a positive dimer-dimer scattering
length $a_{dd}\approx 0.60\, a$. With the quite plausible assumption, that
the total potential energy in a dilute gas of dimers is the sum of its 
{\it two-body} interactions,
the scattering length of the four fermion problem determines the
corresponding interaction constant in the theory of a weakly interacting
Bose gas in the regime of a small gas parameter $n_B^{1/3}a_{dd}\ll 1$, 
where Bogoliubov theory is applicable. The exact dependence 
of the critical temperature of the dilute, repulsive Bose gas on 
the interaction strength has been calculated only in recent  
years. To lowest order in the interaction, the shift is positive and 
linear in the scattering length \cite{Baym99_01},
\begin{equation}
T_c/T_c^{(BEC)}=1+c\,n_B^{1/3}a_{dd}+\ldots
\label{eq:Tc-shift}
\end{equation}
with a numerical constant $c\approx 1.31$ \cite{Arnold01,Kashurnikov01}. 
As a result, the evolution of
the critical temperature in the homogeneous case as a function of the
dimensionless coupling constant $v=1/k_Fa$ necessarily exhibits
a maximum, since the asymptotic ideal Bose gas result is approached
from above. Such a maximum has been found in the early calculations
of $T_c$ along the BCS-BEC crossover by Nozieres and Schmitt-Rink 
\cite{Nozieres} and by Randeria \textit{et al.}\ \cite{Randeria2}. 
The precise height and location of this
maximum, however,  has not been determined so far in a quantitatively 
reliable manner. Given that our present theory exhibits a first 
order transition, there is a range of multi-valuedness of the 
thermodynamic potentials as a function of temperature. 
This regime is bounded in Fig.\ \ref{Fig_phd} by the upper 
and lower $T_c$ curves respectively. 
The lower $T_c$ curve (shown as the red dashed line) which is 
monotonic in $v$ coincides with the $T_c$ curve previously 
calculated \cite{Haussmann2} 
by implementing the Thouless criterion coming from the normal fluid side.
In a situation, where a true first order transition is expected, 
we would need to perform a Maxwell construction to obtain 
the proper transition line. As was discussed above, however,
the first order transition is an artefact of the approximations
involved. In particular the spectrum of excitations right 
at $T_c$ is free particle like in our approximation rather
than $\omega_K\sim K^{3/2}$ \cite{Ferrell67}.

In order to determine the proper critical temperature within our 
approximation , we have used two essentially equivalent criteria: 
the fact that the exact entropy is continuous at $T_c$ suggests 
that our best approximation for the critical temperature is where 
the jump in the entropy between the two branches characterising 
the superfluid and the normal regime has a minimum. Essentially 
the same value is obtained by defining $T_c$ through the criterion 
that it is the maximum temperature at which the order parameter 
$\Delta(T)$ is nonzero. Remarkably, these criteria lead to a 
critical temperature (shown as $\theta_c^{(upper)}$ in  Fig.\ \ref{Fig_phd})
which exhibits a maximum on the BEC-side 
of the crossover around $v \approx 1$ as expected on general 
grounds. Moreover, our theory predicts the correct asymptotic 
functional form \eqref{eq:Tc-shift} of the $T_c$-enhancement 
in the BEC limit $v\gg 1$. Even though the dimer-dimer scattering 
length $a_{dd}^{(B)}=2\, a$ and the prefactor $c\approx 0.58$ of 
our approximate Popov-type theory differ from the exact values
$a_{dd}=0.60\, a$ and $c\approx 1.31$, respectively, the agreement 
of our theory with the exact result is very good (see Fig.\ \ref{Fig_phd}).

A quite sensitive test of the quantitative reliability of our
present result for the critical temperature at arbitrary coupling is 
provided by a comparison with the recent, rather precise numerical
results  right at the unitarity point by Burovski \textit{et al.}\ \cite{Burovski}.
In fact, our result for the dimensionless ratio 
$T_c/\varepsilon_F\approx 0.16$ of the critical temperature 
in units of the bare Fermi energy, which is one of the 
universal numbers of the BCS-BEC crossover problem
(see Subsec.\ \ref{subsection_ul} below), 
agrees precisely with the numerical results of Burovski \textit{et al.}\  
within the given error bars. As will be shown below, a similar
rather precise agreement is obtained with other thermodynamic 
quantities, except for the chemical potential. Thus, even in 
the absence of a small parameter which would allow to 
control our theory systematically in the crossover regime,
the agreement with the numerical results at unitarity
gives us confidence that the approach is quantitatively 
reliable at arbitrary coupling strengths. 

\setlength{\floatsep}{1.0cm}

\begin{figure}
\includegraphics[width=0.4\textwidth]{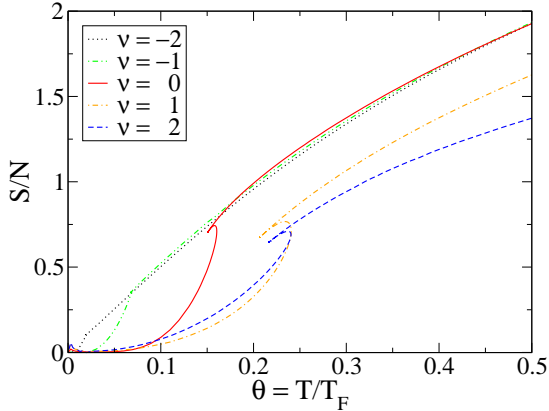}
\caption{(Color online) $S(T)$ at various interaction strengths $v$.}
\label{Fig_S_slices}
\end{figure}

\begin{figure}[t]
\includegraphics[width=0.45\textwidth]{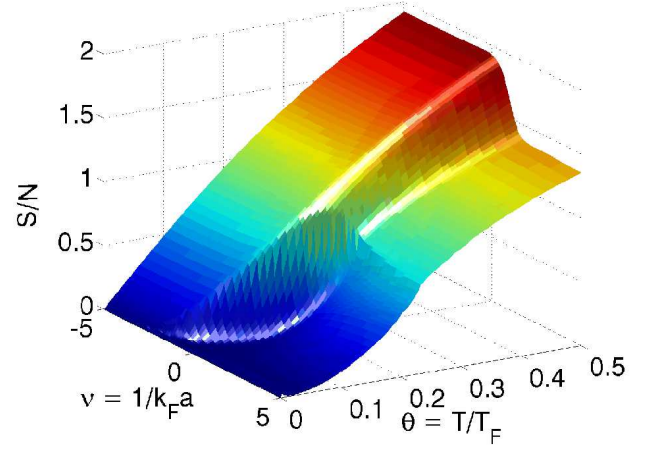}
\caption{Entropy as a function of $\theta$ and $v$ obtained using \eqref{B_510} and \eqref{B_680}.}
\label{Fig_entropy_3d}
\end{figure}

\begin{figure}[t]
\includegraphics[width=0.45\textwidth]{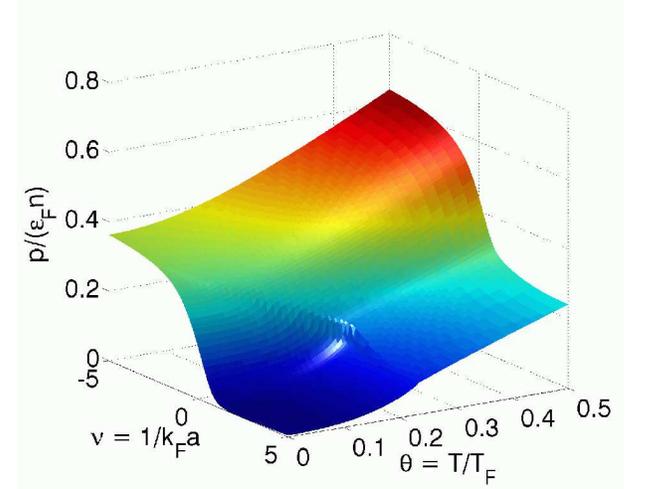}
\caption{Pressure as a function of $\theta$ and $v$ obtained using \eqref{B_490} and \eqref{B_630}.}
\label{Fig_pressure_3d}
\end{figure}

In Fig.\ \ref{Fig_S_slices} the temperature evolution of the entropy is shown 
for various coupling parameters $v$. Here the multivalued character is clearly seen which 
reflects the first-order transition. Furthermore, three-dimensional plots of the entropy 
and of the pressure are presented in Figs.\ \ref{Fig_entropy_3d} and \ref{Fig_pressure_3d}, 
respectively. In both figures a rather sharp drop is observed in the crossover region from weak 
coupling $v \lesssim -1$ (fermionic regime) to strong coupling $v \gtrsim +1$ (bosonic regime).
In the weak coupling limit $v \ll -1 $ the results of the nearly ideal Fermi gas are approached 
which are defined by the BCS formulas \eqref{B_490}-\eqref{B_510} and \eqref{B_620}.
On the other hand, in the strong coupling limit $v \gg +1$ the results of Shohno's mean-field 
theory are approached. While the strong-coupling entropy is defined by \eqref{B_870}, the other
thermodynamic quantities are defined by more complicated formulae \cite{Haussmann4}. For 
$v > 1.0$ and very low temperatures the pressure is nearly zero which reflects a special property of
Shohno's mean-field theory of weakly interacting bosons. At high temperatures $T \gg \varepsilon_F$ 
the entropy, the pressure, and the related thermodynamic quantities approach  
the Boltzmann limit.

\begin{figure}[t]
\includegraphics[width=0.45\textwidth]{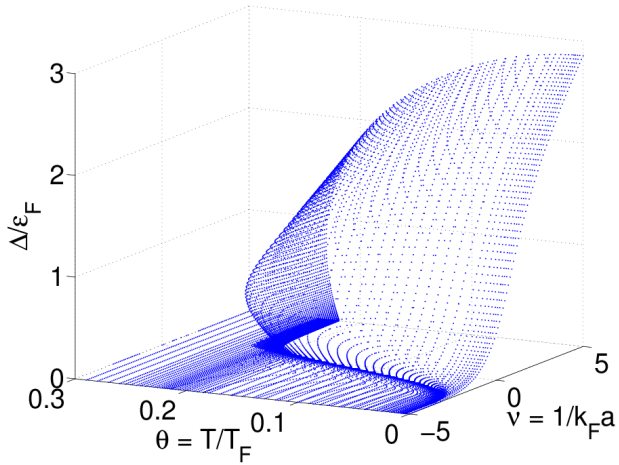}
\caption{(Color online) 3d view of the order parameter.}
\label{Fig_gap_3d}
\end{figure}

Fig.\ \ref{Fig_gap_3d} shows the order parameter which vanishes 
exponentially $\Delta(T=0)/\varepsilon_F \rightarrow (8/e^2) \exp(\pi v/2)$ according to the well 
known BCS result for $v\ll-1.0$. In the opposite limit of strong coupling the behaviour can be 
derived from $\mu n \rightarrow - \Delta^2/2g$ which reflects the fact that the fermion chemical 
potential in the strong coupling limit is governed by the potential (i.e.\ binding) energy. This 
yields $\Delta(T=0)/\varepsilon_F \rightarrow \sqrt{(16/3\pi) v}$ with the square root behaviour 
clearly visible in Fig.\ \ref{Fig_gap_3d}. Near $T_c$ the gap function displays the multivalued 
behaviour characteristic of a first-order transition.    

\begin{figure}
\includegraphics[width=0.45\textwidth]{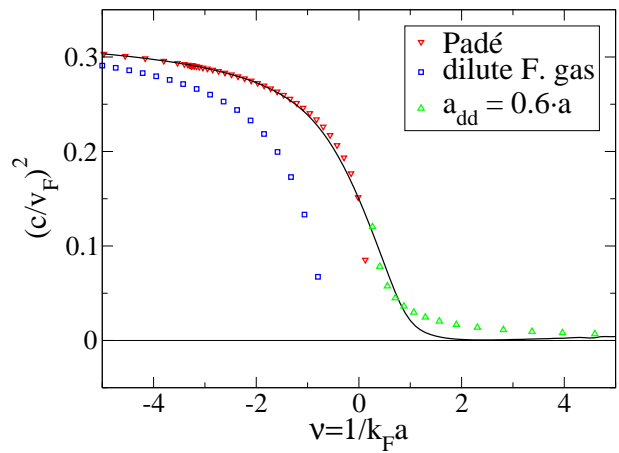}
\caption{(Color online) Isothermal sound speed $mc^2 = \partial p/ \partial n$ as a 
function of $v=1/k_Fa$ for $T=0$. The different curves are explained in the main text.}
\label{Fig_sound_speed}
\end{figure}

At low temperatures the entropy has to vanish, in accordance with the
third law of thermodynamics. The way it does, is in fact universal along the full 
BCS-BEC crossover. Indeed,  at low temperatures, the two-component Fermi gas 
is in a superfluid state, independent of the strength of the attractive  
interaction. On quite general grounds therefore, the low lying excitations above the
ground state are sound modes of the Bogoliubov-Anderson type. These modes give rise 
to an entropy
\begin{equation}
\label{entropy}
S(T)=V \frac{2 \pi^2}{45}\left(\frac{T}{\hbar c}\right)^3+\ldots
\end{equation}
which vanishes like $T^3$ for arbitrary coupling strength. The associated sound 
velocity $c$ is constant at low $T$ and may be determined from the pressure via 
$mc^2=\partial p/\partial n$.
Fig.\ \ref{Fig_sound_speed} displays $(c/v_F)^2$ at $T = 0$ 
as a function of coupling strength with $v_F$ the Fermi velocity.
The dilute interacting Fermi gas limit $(c/v_F)^2=(1+2/(\pi v))/3$
and the BEC limit $(c/v_F)^2=k_F a_{dd}/(6\pi)$ for $a_{dd}=0.60\, a$
are represented by the blue squares and the green triangles respectively. 
The red triangles are obtained by extending the expression of
the ground state energy of a dilute weakly interacting Fermi 
gas \cite{Lenz29,Huang57} with the help of a Pade approximation
to the strong coupling regime \cite{Baker99,Heiselberg01}
\begin{equation}
\frac{E}{\varepsilon_F N} \simeq \frac{3}{5}+\frac{\frac{2}{3\pi}k_Fa}{1-\frac{6}{35\pi}(11-2\ln2)k_Fa}
\end{equation} 
and the thermodynamic identity
\begin{equation}
c^2=\frac{1}{m}\frac{\partial}{\partial n}\left(n^2\frac{\partial E/N}{\partial n} \right) \ .
\end{equation}
Obviously the present crossover theory provides a very good description of the equation
of state and sound velocity except in the regime $v>1$, where our results underestimate both the 
pressure and its density dependence.
 
In principle we should be able to independently obtain $c$ from the low entropy 
asymptotics  \eqref{entropy}. Our numerical results are consistent with 
$S(T)\sim T^3$, however they are not precise enough at such low temperatures, 
to extract the sound velocity in this manner.

\subsection{Thermodynamics in the unitarity limit}
\label{subsection_ul}

After presenting the results for the critical temperature and the thermodynamics at
arbitrary coupling, we now turn to a more detailed discussion of the unitarity limit,
where the scattering length is infinite.
This particular line in the phase diagram has received a lot of 
attention recently. In particular, precise  
numerical results are available at this point \cite{Burovski}, which
provide a sensitive test of analytical approaches
to the crossover problem. 

As has been mentioned before, the Fermi gas at 
infinite scattering length $v=0$ is rather special since
the only relevant length and energy scales remaining in the problem
are  the Fermi wave length set by the density and the Fermi energy
$\varepsilon_F$, provided we remain within the zero range
pseudopotential approximation. The free energy has a simple scaling form
\begin{equation}
F(T,V,N) = f(\theta) N \varepsilon_F \ .
\end{equation}
In particular, there are a number of universal ratios which characterize the
crossover problem right at the unitarity point, both at zero temperature
and at $T_c$.  Examples, which will be determined below, 
are the chemical potential and the internal 
energy in units of the Fermi energy or the entropy per particle at $T_c$. 
In addition, also the gap for single particle excitations or the condensate 
fraction at zero temperature are universal at the unitarity point.

\begin{figure}[t]
\includegraphics[width=0.45\textwidth]{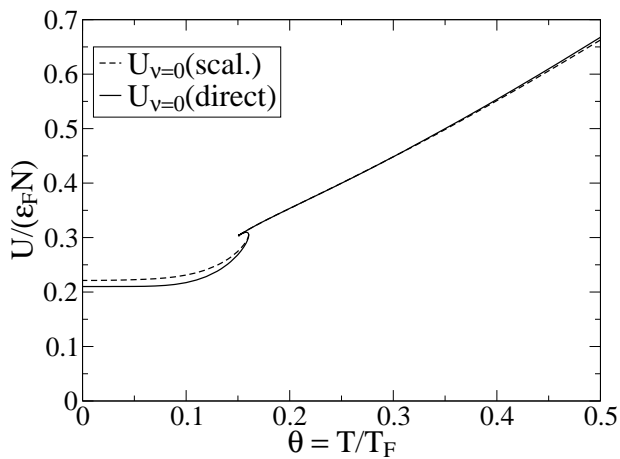}
\caption{Internal energy at unitarity as a function of temperature calculated using
\eqref{B_500} and \eqref{B_670}. The dashed curve is obtained from the calculated
pressure using the scaling formula $U = \frac{3}{2}pV$ valid at unitarity.}
\label{FigULE}
\end{figure}

Fig.\ \ref{FigULE} shows the temperature dependence of the internal
energy calculated in two different ways. The solid  line is our
numerical result   which is compared with the internal energy
(depicted as the dashed line)  as obtained from the numerically
calculated pressure $p = -\Omega/V$ via the scaling relation $U = 3pV/2$ valid at the 
unitarity point. Our numerical results display perfect scaling above $T_c$. The scaling
violation below $T_c$ is a consequence of the modification of the theory. In order to 
preserve the conserving nature of our theory while obeying the Thouless criterion an extra 
length scale $a_\text{mod}$ had to be introduced leading to a modified dimensionless interaction
strength $v_\text{mod} = 1/k_F a_\text{mod}$ with $\delta v_\text{mod} = v_\text{mod} - v$ in 
the range between $0.0$ and $-0.1$ with $v_\text{mod} \neq 0$ for $v = 0$
(see Subsecs.\ \ref{subsection_2I} and \ref{subsection_2J} for details).  
 
\begin{figure}[t]
\includegraphics[width=0.45\textwidth]{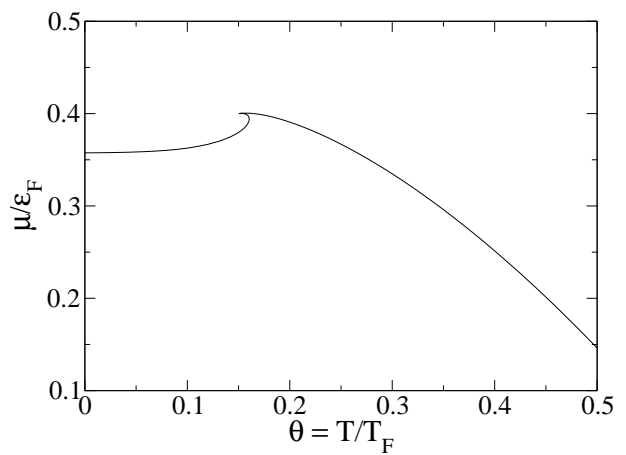}
\caption{The single particle chemical potential at unitarity as a function of temperature 
obtained from the number conservation constraint \eqref{B_220}.}
\label{Figmu}
\end{figure}

Fig.\ \ref{Figmu} displays the behaviour of the chemical potential $\mu(T)$ as 
a function of temperature. Using $\mu(T)$ in a local density approximation
\begin{equation}
\mu = \mu_h\big[n(\textbf{r}),T(\textbf{r})/T_F\big] + V(\textbf{r})
\end{equation}
with $\mu_h$ the chemical potential of the homogenous case we can calculate the density 
profiles of harmonically trapped ultracold gases at unitarity \cite{Rantner06}. We have 
also checked the convergence of our $\mu(T)$ to the high temperature expansion  
obtained by Ho and Mueller \cite{HoMueller04} which however only occurs for $T\gg \varepsilon_F$.  

\begin{figure}
\includegraphics[width=0.45\textwidth]{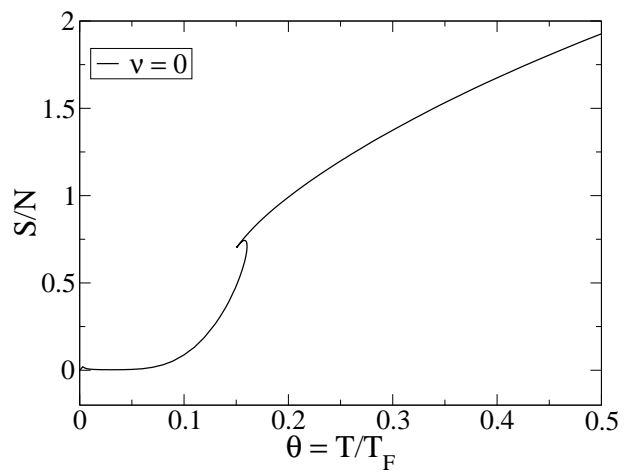}
\caption{Entropy at unitarity as a function of temperature.}
\label{FigULS}
\end{figure}
 
Note that below $T_c$\, the chemical potential $\mu(T)$ is an increasing function of $T$.
This perhaps counterintuitive result can be understood quite easily from the
fact, that the low temperature thermodynamics is determined by the
Bogoliubov-Anderson mode. As argued in the previous section, this leads
to an entropy which vanishes with a power law $\sim T^3$. 
Fig.\ \ref{FigULS} displays the entropy at unitarity as a function of temperature.
Now, at a given volume, there is a Maxwell relation of the form
\begin{equation}
\label{Maxwell}
\frac{\partial\mu}{\partial T}\Big\vert_{N,V}=-\frac{\partial S}{\partial N}\Big\vert_{T,V}
\end{equation}
which connects the temperature dependence of the chemical potential to
the density dependence of the entropy. Using the universal result \eqref{entropy}
for the low temperature entropy, this relation shows that at low
temperatures the chemical potential exhibits a $T^4$ dependence  with a
prefactor determined by
\begin{equation}
\label{dmu/dT}
\frac{\partial\mu}{\partial T}=\frac{3S}{2Vmc_s^2}\frac{\partial^2 p}{\partial n^2}>0 \ .
\end{equation}
Obviously, this argument is not confined to the unitarity point, showing that
the chemical potential at low $T$ has a behaviour $\mu(T)=\mu(0)+\mathcal{O}(T^4)$
for arbitrary coupling strengths along the BCS-BEC crossover.
A well documented quantity which determines the density profile of dilute fermions in 
a trap at unitarity and T = 0 is the so called $\beta$ parameter defined via
\begin{equation}
\label{beta}
\mu(T=0) = \varepsilon_F(1+\beta) \ .
\end{equation}
Our value of $\beta \sim -0.640$ is very close to $\beta= -0.67$ 
obtained via simply Pade approximating the weak coupling
result for the ground state energy \cite{Baker99,Heiselberg01} and the
experimental results of Bartenstein \textit{et al.}\ \cite{Bartenstein04}
$\beta = -0.68^{+0.13}_{-0.10}$ and Bourdel \textit{et al.}\ \cite{Bourdel04} $\beta = -0.64 \pm 0.15$ 
but smaller than the results obtained at Duke \cite{Kinast05}, at Rice \cite{Partridge06} 
and recent QMC results \cite{Carlson03,Astra1} (see Table \ref{Table1}).
Evidently, there is still considerable uncertainty in both the experimental and theoretical results.

\begin{figure}
\includegraphics[width=0.45\textwidth]{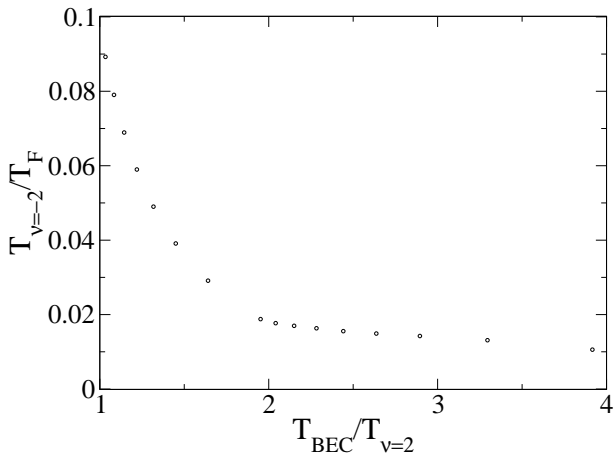}
\caption{Temperature reduction on performing an isentropic sweep across $v=0$ from $v=2$ to $v=-2$.} 
\label{FigIsentrop}
\end{figure}

\begin{table}[t]
\caption{Recent experimental results for $\beta$ compared with calculated
values.}
\begin{tabular}{l | l | l}
\multicolumn{3}{l}{} \\
             &                                                     & \phantom{cent}$\beta$   \\
\hline
             & Bartenstein \textit{et al.}\ \cite{Bartenstein04}   & $-0.68^{+0.13}_{-0.10}$ \\
Experimental & Bourdel (2004) \textit{et al.}\ \cite{Bourdel04}    & $-0.64(15)$             \\
results      & Duke (2005) \cite{Kinast05}                         & $-0.49(4)$              \\
             & Partridge \textit{et al.}\ \cite{Partridge06}       & $-0.54(5)$              \\
\hline \hline
                & Astrakharchik \textit{et al.}\ \cite{Astra1}     & $-0.58(1)$              \\
                & Carlson \textit{et al.}\ \cite{Carlson03}        & $-0.56(1)$              \\
Calculated      & Hu/Liu/Drummond \cite{Hu_europhys}               & $-0.599$                \\
values          & Perali \textit{et al.}\ \cite{Perali93}          & $-0.545  $              \\
                & Pad\'e approximation \cite{Baker99,Heiselberg01} & $-0.67$                 \\
                & present work                                     & $-0.64   $              \\
\end{tabular}
\label{Table1}
\end{table}

A promising route in the direction of thermometry for trapped gases is provided via the
reversible adiabatic (isentropic) sweeps \cite{Bartenstein04,Chin}
from the BEC limit. In Fig.\ \ref{FigIsentrop} we depict the
resulting changes in temperature when moving across the unitarity
limit for the homogenous case. For the trapped case this cooling
mechanism was first advocated by Carr \textit{et al.}\ \cite{Carr} and
recently quantitatively refined by Hu \textit{et al.}\ \cite{Hu}.  
Finally to facilitate quantitative comparison with various Quantum
Monte Carlo results we have collected available data from the
literature presented in  Table \ref{Table1} and \ref{Table2}. 
Apart from the value for $T_c$ which is explicitely quoted in the
paper by Bulgac (with errors) we have estimated the remaining
quantities from their presented results and utilized scaling to fill
in the missing data below.

\begin{table}
\caption{Comparison with diagrammatic determinant Mon\-te Carlo 
(Burovski \textit{et al.}\ \cite{Burovski}), quantum Monte Carlo 
(Bulgac \textit{et al.}\ \cite{Bulgac}), $\varepsilon=4-d$ expansion
(Nishida and Son \cite{Nishida15_08_06,Nishida31_07_06}), Borel-Pad\'e approximation connecting 
an expansion in $\varepsilon=4-d$ and one in $\varepsilon=d-2$ 
\cite{Nishida15_08_06}) and a $1/N$ expansion (Nikoli\'c and Sachdev)\cite{Nikolic} at $T=T_c$.}
\begin{tabular}{l| l | l | l | l | l}
\multicolumn{6}{l}{} \\
 &$T_c/\varepsilon_F$ & $\mu/\varepsilon_F$ & $U/N\varepsilon_F$ &
 $P/n \varepsilon_F$ & $S/N$ \\
\hline
Bulgac & 0.23(2) & 0.45 &0.41 & 0.27 & 0.99 \\
Burovski & 0.152(7) & 0.493(14) & 0.31(1) & 0.207(7) & 0.16(2) \\
Nikoli\'c {\scriptsize ($N=1$)}& 0.136 & 0.585 &0.164 & 0.109 & \\
Nishida {\scriptsize ($\varepsilon=1$)} & 0.249 & 0.18 & 0.212 & 0.135 & 0.698 \\ 
Borel-Pad\'e & 0.183 & 0.294 & 0.270 & 0.172 & 0.642 \\
present work & 0.160 & 0.394 & 0.304 & 0.204 & 0.71 \\
\end{tabular}
\label{Table2}
\end{table} 

\begin{table}
\caption{Comparison with fixed node Green function Monte Carlo 
(Astrakharchik \textit{et al.}\ \cite{Astra1} and Carlson \textit{et al.}\ \cite{Carlson03}) 
at $T=0$}
\begin{tabular}{l | l | l | l |l}
\multicolumn{5}{l}{} \\
 & $\mu/\varepsilon_F$ & $U/N\varepsilon_F$ & $P/n \varepsilon_F$ & $\Delta/\varepsilon_F$ \\
\hline
Astrakharchik  &0.41(2)  & 0.25(1) & 0.17(1) & \\
Carlson  &0.43(1) & 0.26(1) & 0.17(1) & 0.54 \\
present work & 0.36 & 0.21 & 0.15 & 0.46 \\
\end{tabular}
\label{Table3}
\end{table}

The $T=0$ results are fixed node QM results by Astrakharchik 
\textit{et al.}\ \cite{Astra1} and Carlson \textit{et al.}\ \cite{Carlson03}. 
Note that our result for $\Delta/\varepsilon_F$ is close to the value 
$\Delta_\text{GMB}/\varepsilon_F=(2/e)^{7/3}=0.49$ obtained by a naive extrapolation 
of the Gorkov Melik-Barkudarov result to $k_F a= \infty$.

At $T_c$  our results are in very good agreement with those
of Burovski \textit{et al.}\ except  for the value of the dimensionless chemical
potential $\mu/\varepsilon_F$ and that of  the entropy per particle 
at $T_c$. Now Burovski \textit{et al.}\ have obtained their values
for the pressure $p/n \varepsilon_F$ and the entropy $S/N $ 
indirectly from the internal energy and the chemical potential by 
using $3pV=2U$ right at unitarity and the Gibbs-Duhem relation.
The different results for the chemical potential then entail the
considerable discrepancy in the value of $S/N$ at $T_c$.
Within our numerical scheme, the chemical potential is the most
directly - via \eqref{B_220} - obtainable quantity among the
thermodynamic data. In light of the excellent agreement  
of all other quantities with the numerical results of Burovski \textit{et al.},
the discrepancy for the chemical potential is thus quite surprising.
Indeed, we believe that our values for both the chemical potential
and the entropy, for which the validity of the Gibbs-Duhem relation and 
of $3pV=2U$ at unitarity have been checked {\it independently},
are rather close to the exact results. This point of view is
supported by considering the evolution of the entropy per particle 
right at the critical temperature as a function of the dimensionless
coupling.  In the BCS limit, the entropy associated with single 
particle excitations can be calculated from the exactly soluble
reduced BCS-Hamiltonian and is given by the standard mean-field expression
\eqref{B_510}. At the critical temperature this entropy coincides with that 
of an ideal Fermi gas
\begin{equation}
\label{BCS-entropy}
S(T_c)/N= (\pi^2/2) (T_c/T_F) \ .
\end{equation}

Since the ratio $T_c/T_F$ is exponentially small in the weak
coupling limit, the entropy \eqref{BCS-entropy} associated with 
fermionic excitations is dominant compared to the contribution
arising from the collective Bogoliubov-Anderson mode. Indeed,
extrapolating the corresponding low temperature entropy \eqref{entropy} 
associated with collective excitations up to the critical temperature
gives rise to a contribution of order $(T_c/T_F)^3$, which is 
negligible compared to \eqref{BCS-entropy}. 

\begin{figure}[t]
\includegraphics[width=0.4\textwidth]{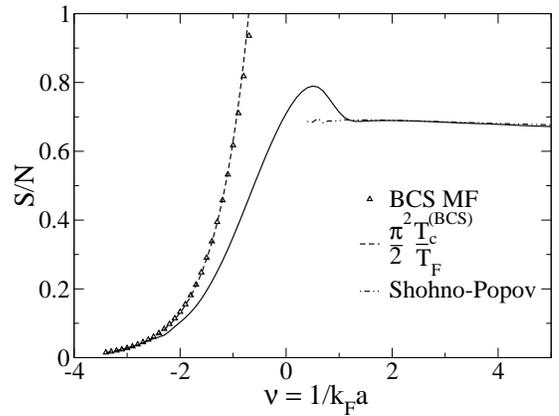}
\caption{Entropy at $T_c$ as a function of $v = 1/k_Fa$. Numerical result
(solid) line obtained with \eqref{B_510} and \eqref{B_680} compared 
with the limiting results: the BCS mean-field result (triangles) from \eqref{B_510}
and (dashed line) from \eqref{BCS-entropy}
and the Shohno-Popov result (dotted-dashed line) from \eqref{B_880}.}
\label{Fig_utc_entropy}
\end{figure}

At very large coupling 
strengths, the strongly bound fermion pairs form an eventually ideal
Bose gas, for which the entropy per particle right at $T_c$ can 
again be determined analytically. Recalling, that the number 
of bosons $N_B=N/2$ in this limit is just half the number of fermions,
we obtain a universal number 
\begin{equation}
\label{BEC-entropy}
S(T_c)/N=\frac{5\zeta(5/2)}{4\zeta(3/2)}=0.6417... \, .
\end{equation}
As is evident from Fig.\ \ref{Fig_utc_entropy}, where the complete evolution of the 
ratio $S(T_c)/N$ is shown as a function
of the dimensionless coupling parameter $v$, 
the limiting value of the ideal Bose gas is in fact not far 
from the entropy which is obtained from the Shohno-Popov
theory of noninteracting bosonic quasiparticles in the 
range $v>1$, according to \eqref{B_870}.

It is interesting to note, that the entropy per particle right at $T_c$
exhibits a maximum as a function of the coupling constant of order
$S(T_c)/N\approx 0.78$ around the same coupling, where
the critical temperature exhibits a maximum. Considering the smooth evolution
of $S(T_c)/N$ as a function of $v$, the value $S(T_c)/N\approx 0.16$
at unitarity, which is deduced from the results of Burovski \textit{et al.}, appears
to be far too small. On the other hand, the result $S(T_c)/N\approx 0.99$ obtained 
by Bulgac \textit{et al.}\ seems to be too high.

\section{Discussion and Conclusion}
\label{Discussion}

In conclusion let us summarize what has been achieved, mention shortcomings of
the present approach and indicate possible future extensions.

The formal basis of our results is a self-consistent, conserving theory, which is based 
on an approach due to Luttinger-Ward and DeDominicis-Martin, in which the 
exact one- or two-particle Green functions serve as an infinite set of 
variational parameters. In order for this approach to provide
consistent thermodynamic results it is essential that the Green functions satisfy 
self-consistency conditions which reflect the stationarity of the appropriate thermodynamic 
potentials. Approximate formulations, in which free Green functions are replaced by full 
ones according to a choice of $G_0G_0$, $GG_0$ or $GG$ will in general not obey conservation 
laws or exact thermodynamic identities, in contrast to the $\Phi$ derivable formulation 
presented here. The stationarity conditions were also crucial for the proof of thermodynamic 
equivalence of the Luttinger-Ward with the DeDominicis-Martin formalism on the level of our 
approximate functional for the grand canonical potential or the entropy. In fact, to
our knowledge, the theory presented here is the first concrete application of the 
DeDominicis-Martin formulation to the fermionic many-body problem.

An important point, we want to emphasize, is the necessarily 
self-consistent nature
of the formalism. Indeed, within the Luttinger-Ward or the
DeDominicis-Martin formulation an approximate functional
for the grand canonical potential $\Omega[G]$ or the
entropy $S[G,\Gamma]$ is made stationary by
determining the space- and time-dependent Green and
vertex functions from the variational conditions
\eqref{B_120} and \eqref{B_126} respectively. The solution of these
equations {\it necessarily} leads to a self-consistent
mutual dependence of the various Green functions.
Self consistency is thus reached precisely at the stationary
point of these functionals. At this point, equations \eqref{B_120} and \eqref{B_126}
are valid, conditions which are necessary  for the
theory to give consistent thermodynamics, as pointed out
e.g.\ in the context of equation \eqref{B_260}.

A well known shortcoming of conserving approximations is the dichotomy with the gapless 
nature of the collective modes, which reflects the broken continuous symmetry of the 
superfluid state. For the present theory the formal reason for this dichotomy is a 
violation of the Ward identity resulting from the global gauge symmetry of the exact 
theory. In order to overcome this problem an extension of the theory was introduced 
which forces the gapless nature in the symmetry broken phase while remaining $\Phi$-derivable 
at the same time to maintain the conserving property.

We have provided quantitative results for essentially all
thermodynamic properties at temperatures below half the
Fermi temperature, thus covering the relevant regime of the
degenerate gas. Overall our results agree remarkably well with
recent numerical calculations at the unitarity point giving
confidence that our approach is quantitatively reliable over
the full range of couplings between the BCS- and the BEC-limit.
In particular, we provide concrete predictions for a number of
universal ratios characterizing the unitary Fermi gas both
at $T=0$ and at $T=T_c$.

The extensive numerical work entering the solution of the stationarity constraints and 
thermodynamic potentials is reflected most clearly in the three dimensional plots of the 
entropy Fig.\ \ref{Fig_entropy_3d}, pressure Fig.\ \ref{Fig_pressure_3d} and the order 
parameter Fig.\ \ref{Fig_gap_3d}. Most noteworthy are the quite abrupt change from fermionic 
to bosonic character for $v$ in the
interval $-1<v<+1$ which are most obvious in the entropy and pressure and the quick 
convergence to a Shohno-Popov theory of noninteracting bosonic quasiparticles for $v > 1$. 

An initially unexpected result,  which is clearly visible in the numerical data, is the 
fact that our superfluid phase transition is weakly first order, instead of being continuous
as it should be. The origin of this failure to capture the critical behaviour correctly is 
found in the Shohno-Popov theory, which is obtained from our approach in the limit $v\gg 1$. 
The Shohno-Popov theory of a dilute, repulsive Bose gas generalizes the Bogoliubov theory
to finite temperatures. It takes into account the thermal depletion of the condensate 
by including the effect of bosons with finite momentum  $\textbf{K}$ in the particle 
number equation.  
Long ago Reatto and Straley \cite{ReattoStraley69} analyzed Shohno's theory in 
a self-consistent formulation and obtained a first-order superfluid transition. 
Physically, the origin of the associated entropy jump is
the collapse of the single particle spectrum right at the transition. Indeed, within the 
Shohno-Popov theory,  the single particle spectrum changes  from 
initially linear to initially quadratic on raising the temperature through $T_c$. 
As a result, the density of states is changed from a $\varepsilon^2$ dependence 
below $T_c$  to the free particle $\sqrt{\varepsilon}$ result right at and above
$T_c$. The associated drastic increase in the available phasespace 
leads to a jump in the entropy. 

For a purely bosonic system, a proper treatment of the behaviour near the critical 
point was recently given by Baym and coworkers \cite{Baym99_01}.
Baym and Holzmann \cite{Baym03} showed that a
change of the spectrum for long wavelength excitations occurs right at
$T_c$. This hardening of the spectrum (the low $\textbf{K}$ spectrum is of the
form $\textbf{K}^\alpha$ with $\alpha < 2$) leads to the required reduction in the 
density of states to render the superfluid
transition continuous. The subtle low $\textbf{K}$ correlations necessary for this 
change in spectrum are clearly missing in our self-consistent approach. 

The BCS-BEC crossover being continuous however implies that the first order result 
also pertains to the $v\ll-1$ limit of our theory. We have checked that at the 
transition the discontinuities of all thermodynamic quantities are 
$\sim \exp( -C \vert v \vert )$ for $v\ll-1$ \cite{Haussmann4}. The associated 
difficulties of a proper treatment of bosonic excitations do not occur in the 
reduced BCS hamiltonian which neglects collective modes altogether, resulting in 
a continuous superfluid transition.
To correctly account for the critical regime $\Delta T / T_c \rightarrow 0$ 
our theory would need to be extended to treat the feedback between different bosonic 
modes accurately.
Bickers and Scalapino \cite{BS92} have shown that this requires the incorporation of single particle self consistency and two
particle self consistency on the same level of approximation.
This may be achieved via so called parquet resummations. 
Currently, however, the inclusion of these contributions appears extremely challenging. 
A systematic and analytically accessible description of the crossover 
which is uniformly valid in both the normal and superfluid regime and which gives a proper account
of the critical behaviour is provided by a $1/N$-expansion as recently shown by 
Nikoli\'c and Sachdev \cite{Nikolic}. This method can in fact be extended in a straigthforward manner
to the case of unbalanced spin populations, a subject which has attracted a lot of attention 
very recently \cite{Zwierlein06,Partridge06}.

\acknowledgments
\noindent
We acknowledge useful discussions with J.N.\ Fuchs, R.\ Grimm and 
M.\ Zwierlein and, moreover, would like to acknowledge
an illuminating exchange with Y. Nishida
and S. Sachdev on their recent work.
Part of this work was supported by the Deutsche Forschungsgemeinschaft DFG 
within the Schwerpunkt ``Ultrakalte Gase'' (S.\ Cerrito).

\appendix

\section{Regularization of divergent Matsubara-frequency sums} 
\label{appendix_B}

In our formulas of the thermodynamic potentials most sums over Matsubara frequencies
are not well defined. The functions which are summed do not decay to zero fast enough
so that the Matsubara-frequency sums diverge. However, this problem can be fixed. To
do this we first perform a Fourier back transformation to obtain a function in terms
of the imaginary time $\tau$. Then we take the limit $\tau\rightarrow -0$ or 
$\tau\rightarrow +0$ which is finite and well defined.

We must distinguish between fermion and boson functions which have different 
Matsubara frequencies. Fermion functions are of the type
\begin{equation}
A(\textbf{k},\omega_n) =
\begin{pmatrix}
{\cal A}(\textbf{k},\omega_n) &{\cal B}(\textbf{k},\omega_n)\cr
-{\cal B}(\textbf{k},\omega_n)^* &{\cal A}(\textbf{k},\omega_n)^* 
\end{pmatrix}
\label{X_110}
\end{equation}
where $A(\textbf{k},\omega_n)$ may be either $A(\textbf{k},\omega_n)=-\ln [G(\textbf{k},\omega_n)]$ \newline
or $A(\textbf{k},\omega_n)= [G_0(\textbf{k},\omega_n)^{-1} G(\textbf{k},\omega_n) - 1]$. 
(Note that the lower row of the matrix \eqref{X_110} has the opposite sign than the lower
row of the matrix \eqref{B_140}. The reason ist that in the terms of the thermodynamic 
potentials always an even number of fermion Green functions is multiplied together.)
In this case we define
\begin{equation}
\frac{1}{\beta} \sum_{\omega_n} \text{Tr}\{ A(\textbf{k},\omega_n) \} =
2\, {\cal A}(\textbf{k},\tau=-0) 
\label{X_120}
\end{equation}
where we assume that ${\cal A}(\textbf{k},\tau)$ is real. Similarly consider a bosonic 
function of the form 
\begin{equation}
A(\textbf{K},\Omega_n) =
\begin{pmatrix}
{\cal A}(\textbf{K},\Omega_n) &{\cal B}(\textbf{K},\Omega_n)\cr
{\cal B}(\textbf{K},\Omega_n)^* &{\cal A}(\textbf{K},\Omega_n)^* 
\end{pmatrix}
\label{X_130}
\end{equation}
with $A(\textbf{K},\Omega_n)=\Gamma(\textbf{K},\Omega_n)$ \newline 
or $A(\textbf{K},\Omega_n)=-\ln [\Gamma(\textbf{K},\Omega_n)]$. In this case we define
\begin{equation}
\frac{1}{\beta} \sum_{\Omega_n} \text{Tr}\{ A(\textbf{K},\omega_n) \} =
2\, {\cal A}(\textbf{K},\tau=-0) \ ,
\label{X_140}
\end{equation}
where we assume that ${\cal A}(\textbf{K},\tau)$ is real.

In some terms of our formulas the fermion function $A(\textbf{k},\omega_n)$ or the 
boson function $A(\textbf{K},\Omega_n)$ is proportional to the unit matrix $1$. 
In this case the Fourier backtransform ${\cal A}(\textbf{K},\tau)$ is 
$\delta_F(\tau/\hbar)$ or $\delta_B(\tau/\hbar)$, respectively. Hence, the related 
Matsubara-frequency sums \eqref{X_120} or \eqref{X_140} are zero.

\section{Numerical Fourier transformation} 
\label{appendix_D}

The special numerical Fourier transformation has been invented long time ago by one of 
the authors in a different context in order to solve the mode-coupling equation for the 
liquid-glass transition \cite{GH88}. In this case relaxation phenomena are considered on 
a logarithmic time scale over many decades, starting at microscopically short times and 
extending up to very long macroscopic times. Thus, a Fourier transformation 
is needed which can handle functions with features on logarithmic time and frequency 
scales extending over ten and more decades. Clearly, a standard fast Fourier 
transformation can not be applied because a constant step width would be needed. 
Rather the function to be transformed has been discretized on a logarithmic scale and 
interpolated by cubic spline polynomials. Since for polynomial functions the Fourier 
integrals can be evaluated exactly, we end up with a transformation formula which 
depends on the spline coefficients of the function.

Later this special numerical Fourier transformation has been extended to transform 
Matsubara Green functions in order to solve the self-consistent equations for
the BCS-BEC crossover \cite{Haussmann2}. Here, three-dimensional spatial Fourier 
transformations of isotropic functions and discrete Fourier sums with Matsubara
frequancies were considered. These Fourier transformations are used also in the present
paper for the numerical calculations. Only a few modifications and optimizations have
been made over the years. The basic principles of the special numerical Fourier 
transformation are descibed in the appendix of Ref.\ \onlinecite{Haussmann2}. Here 
we present the fundamental formulas in order to make the numerical method available 
for applications.

In order to perform a discrete Fourier transformation the following sum must be evaluated
\begin{equation}
f(k) = \sum_{x=x_\text{min}}^{x_\text{max}} \Delta x \ e^{ikx} \, f(x )
\label{X_310}
\end{equation}
where $x$ is a discrete variable with constant step width $\Delta x$. In this formula and
in the following formulas the sum over $x$ is defined as a trapezoid sum. This means that 
the first term and the last term in the sum are multiplied by a factor $\frac{1}{2}$, 
respectively. The continuous Fourier transformation is defined by a related integral which 
is obtained from \eqref{X_310} in the limit $\Delta x \rightarrow 0$.

We assume that the function values are known in a finite subset of points $x_j$ according to
$f(x_j) = a_j$ where $j=0,1,\ldots,N$. The points $x_j$ cover the whole interval between
$x_\text{min}$ and $x_\text{max}$ on a logarithmic scale so that 
$x_\text{min}=x_0<x_1<\ldots <x_{N-1}<x_N=x_\text{max}$. Consequently, the Fourier sum
\eqref{X_310} can be divided into a sum of $N$ trapezoid sums according to
\begin{equation}
f(k) = \sum_{j=0}^{N-1} \Bigl\{ \sum_{x=x_j}^{x_{j+1}} \Delta x \ e^{ikx} \, f(x ) \Bigr\} \ .
\label{X_320}
\end{equation}
Now, we assume that the function is given by the cubic spline polynomial
\begin{equation}
f(x) = a_j + b_j(x-x_j) + c_j(x-x_j)^2 + d_j(x-x_j)^3
\label{X_330}
\end{equation}
if $x$ is located in the interval $x_j \leq x \leq x_{j+1}$. The spline coefficients $a_j$,
$b_j$, $c_j$, and $d_j$ are calculated numerically. Inserting the cubic spline polynomial 
\eqref{X_330} into the formula \eqref{X_320} we find that the trapezoid sums within the curved
brackets can be evaluated exactly. Thus, as a result we obtain the Fourier transform
\begin{equation}
f(k) = \sum_{j=0}^{N-1} \Bigl\{ a_j I_j^{(0)}(k) + b_j I_j^{(1)}(k) + c_j I_j^{(2)}(k) 
+ d_j I_j^{(3)}(k) \Bigr\}
\label{X_340}
\end{equation}
where
\begin{equation}
I_j^{(n)}(k) = e^{ikx_j} \Bigl( -i \frac{\partial}{\partial k} \Bigr)^n \Bigl[ \frac{\Delta x}{2i} 
\cot\Bigl(\frac{k \Delta x}{2} \Bigr) \bigl[ e^{ik(x_{j+1}-x_j)} - 1 \bigr] \Bigr] \ .
\label{X_350}
\end{equation}

By construction a cubic spline function and its first two derivatives are continuous. These
facts imply the following continuity conditions
\begin{widetext}
\begin{eqnarray}
f(x_{j+1}) &=& a_j + b_j(x_{j+1}-x_j) + c_j(x_{j+1}-x_j)^2 + d_j(x_{j+1}-x_j)^3 = a_{j+1} \ , \\
\label{X_360}
f^\prime(x_{j+1}) &=& b_j + 2 c_j(x_{j+1}-x_j) + 3 d_j(x_{j+1}-x_j)^2 = b_{j+1} \ , \\
\label{X_370}
f^{\prime\prime}(x_{j+1}) &=& 2 c_j + 6 d_j(x_{j+1}-x_j) = 2 c_{j+1} \ ,
\label{X_380}
\end{eqnarray}
\end{widetext}
which may be used to regroup the terms in \eqref{X_340}. Consequently, as a result we obtain
the alternative formula
\begin{equation}
\begin{split}
f(k) =& J^{(0)}(k) \bigl[ e^{ik x_N} a_N - e^{ikx_0} a_0 \bigr] \\
&+ J^{(1)}(k) \bigl[ e^{ik x_N} b_N - e^{ikx_0} b_0 \bigr] \\
&+ J^{(2)}(k) \bigl[ e^{ik x_N} c_N - e^{ikx_0} c_0 \bigr] \\
&+ J^{(3)}(k) \sum_{j=0}^{N-1} \bigl[ ( e^{ikx_{j+1}} - e^{ikx_j} ) d_j \bigr]
\end{split}
\label{X_390}
\end{equation}
where
\begin{equation}
J^{(n)}(k) = \Bigl( -i \frac{\partial}{\partial k} \Bigr)^n \Bigl[ \frac{\Delta x}{2i} 
\cot\Bigl(\frac{k \Delta x}{2} \Bigr)  \Bigr] \ .
\label{X_400}
\end{equation}
The terms with spline coefficients $a_j$, $b_j$, and $c_j$ have cancelled for $j=1,2,\ldots,N-1$.
In the limit $k\rightarrow 0$ the functions \eqref{X_400} diverge according to 
$J^{(n)}(k)~\sim \vert k\vert^{-(n+1)}$. For this reason, the alternative formula \eqref{X_390} 
can be applied numerically only for large $k$ so that $\vert k x_j\vert \gtrsim 1$ for all 
$j=0,1,\ldots,N$. On the other hand, the functions \eqref{X_350} are finite in the limit 
$k\rightarrow 0$ so that the formula \eqref{X_340} can be applied numerically for small $k$ 
where $\vert k x_j \vert\lesssim 1$ for all $j=0,1,\ldots,N$. In practice we use a combination 
of both formulas \eqref{X_340} and \eqref{X_390}. Which formula is used for a particular $j$ we 
decide by considering the value of $\vert k x_j\vert$ and comparing this value with $1$. In this 
way we obtain a special numerical Fourier transformation which is stable and reliable 
for points $x_j$ and $k_l$ distributed on a logarithmic scale over many decades.

We have derived our special numerical Fourier transformation for discrete variables $x$ with
a finite constant step width $\Delta x$. The continuous Fourier transformation is obtained
easily and naturally by taking the limit $\Delta x \rightarrow 0$ in the functions \eqref{X_350}
and \eqref{X_400} which is well defined.

In order to transform the Green and vertex functions forward and backward, we need two kinds 
of Fourier transformations. First we transform between the Matsubara frequencies and the imaginary
time variable. In this case we can apply a continuous (forward) and a discrete (backward) 
Fourier transformation \eqref{X_310} directly. Secondly we transfrom between the wave vector and
the spatial coordinate in $d=3$ dimensions. Since the functions are spherically symmetric, an
integration over the angles can be performed, so that the resulting transformation integrals
are one dimensional depending only on radial variables, a radial wave number and a radial space
coordinate, respectively. For $d=3$ the transformation integrals can be recast into a 
one-dimensional continuous Fourier transformation so that our special numerical Fourier 
transformation \eqref{X_310} can be used once again.

In practice we use $N=300$ points for all variables. The values of the wave numbers and the values 
of the space coordinates are distributed on logarithmic scales over six decades, respectively.
The Matsubara frequencies are distributed on a logarithmic scale over about twelve decades.
The imaginary time variables are distributed appropriately over a finite interval with two
logarithmic scales, one for each boundary.

The Green and vertex functions are singular and have slowly decaying long tails. For this reason,
reference functions must be subtracted which remove the singularities and the long tails. The
reference functions are derived from free Green functions and the two-particle scattering amplitude
($T$ matrix). For these reference functions analytical expressions must be available in all
Fourier representations. The difference functions $f(x)$ which are eventually transformed by
our numerical method \eqref{X_310} must be smooth in $x$ and decay accoording to $f(x)\sim x^{-2}$
or faster for $\vert x\vert \rightarrow \infty$.

\end{document}